\documentclass[twocolumn,showpacs,aps,prc,superscriptaddress,nofootinbib]{revtex4-1}
\usepackage{atlasphysics} 
\usepackage{subfigure}
\usepackage{mathrsfs}
\usepackage{graphicx}
\usepackage{multirow}
\usepackage{rotating}
\usepackage{xspace}
\usepackage{preprintcover}
\usepackage{color}
\usepackage{amsmath}
\usepackage{morefloats}
\usepackage{placeins}
\usepackage{graphicx}
\usepackage[breaklinks,pdftex,colorlinks]{hyperref}
\hypersetup{linkcolor=black,citecolor=blue,filecolor=black,urlcolor=blue} 
\def\papertitle{
Measurement of differential $\jpsi$ production cross-sections and forward-backward ratio in {\em p}+Pb collisions with the ATLAS detector
}

\usepackage{preprintcover}

\PreprintCoverPaperTitle{\papertitle}
\PreprintIdNumber{CERN-PH-EP-2015-103}

\PreprintJournalName{Physical Review C}

\def\MyAbstract{
Measurements of differential cross-sections for $\jpsi$ production in {\em p}+Pb collisions at $\sqrt{s_{NN}}$ = 5.02$\TeV$ at the LHC with the ATLAS detector are presented. The data set used corresponds to an integrated luminosity of 28.1 $\mathrm{nb}^{-1}$. The $\jpsi$ mesons are reconstructed in the dimuon decay channel over the transverse momentum range $8<\pt<30\GeV$ and over the center-of-mass rapidity range $-2.87<y^{*}<1.94$. Prompt $\jpsi$ are separated from $\jpsi$ resulting from $b$-hadron decays through an analysis of the distance between the $\jpsi$ decay vertex and the event primary vertex. The differential cross-section for production of nonprompt $\jpsi$ is compared to a FONLL calculation that does not include nuclear effects. Forward-backward production ratios are presented and compared to theoretical predictions. These results constrain the kinematic dependence of nuclear modifications of charmonium and $b$-quark production in {\em p}+Pb collisions.
}

\PreprintCoverAbstract{\MyAbstract}

\begin{document}
\title{\papertitle}
\author{The ATLAS Collaboration}

\begin{abstract}
\MyAbstract
\end{abstract}

\pacs{25.75.Cj}

\maketitle

\section{Introduction}
Quarkonium production in heavy-ion collisions is expected to be highly sensitive to the nature of the hot and dense matter created in these collisions~\cite{Brambilla:2010cs}. Suppression of the $\jpsi$ yield in nucleus-nucleus ($A$+$A$) collisions with respect to proton-proton ($pp$) collisions was predicted to be a signal for deconfinement in the quark-gluon plasma~\cite{Matsui:1986dk}. Such suppression was observed at fixed-target experiments at the SPS~\cite{Abreu:1997jh,Abreu:2000ni,Alessandro:2004ap,Adare:2006ns,Arnaldi:2007zz} and in collider experiments at RHIC~\cite{Adare:2011yf,Abelev:2009qaa,Adamczyk:2013tvk} and the LHC~\cite{Aad:2010aa,Abelev:2012rv,Chatrchyan:2012np}. The interpretation of these results is complicated by the fact that the suppression was also observed in proton-nucleus ({\em p}+$A$)~\cite{Abelev:2013yxa,Aaij:2013zxa,Badier:1983dg,Leitch:1999ea,Abt:2008ya,Arnaldi:2010ky} and deuteron-nucleus ($d$+$A$)~\cite{Adare:2010fn} collisions, where final-state effects due to hot matter are not expected. 

Several phenomenological interpretations have been proposed to explain the suppression observed in {\em p}+$A$ or $d$+$A$ collisions.  These include nuclear absorption~\cite{Capella:1988ha,Kharzeev:1995br,Kharzeev:1996yx,Vogt:1999cu}, modifications of parton distribution functions in nuclei (shadowing)~\cite{Kopeliovich:2001ee,Vogt:2004dh,Kopeliovich:2002yv,Ferreiro:2013pua,Ferreiro:2009ur}, gluon saturation~\cite{Kharzeev:2008nw,Kopeliovich:2010aa,Fujii:2013gxa,Kang:2013hta,Kharzeev:2005zr}, and in-medium energy loss~\cite{Gavin:1991qk,Arleo:2012rs}. For a review of these cold-medium effects see Ref.~\cite{ConesadelValle:2011fw}. 
The impact of each of these mechanisms on $\jpsi$ production varies with rapidity and transverse momentum. Measurements at large rapidities probe the low-$x$ partons in the nuclei, and gluon shadowing and saturation effects are expected to be important. 

The cold-medium processes that affect quarkonia production can also affect $b$-quark production. The effects of gluon saturation and shadowing are expected to be similar to those for charmonium production, but nuclear absorption and parton energy loss are expected to be less pronounced. Therefore, additional constraints can be obtained by measuring $b$-quark production, which can be accomplished by measuring the cross-section for $\jpsi$ production in the decay chains of $b$-hadrons; these are abbreviated as ``nonprompt $\jpsi$.'' 
  
Measurements in {\em p}+$A$~\cite{Leitch:1999ea,Abt:2008ya,Arnaldi:2010ky,Abelev:2013yxa,Aaij:2013zxa} and $d$+$A$~\cite{Adare:2010fn} collisions show that the differential cross-section for $\jpsi$ production as a function of the center-of-mass rapidity\footnote{The center-of-mass rapidity is defined as $y^{*}=\frac{1}{2} \ln{\left( \frac{E+p_{z}}{E-p_{z}} \right)}$, where $E$ and $p_{z}$ are the energy and the component of the momentum along the proton beam direction in the nucleon-nucleon center-of-mass frame.} $y^{*}$ is not symmetric around $y^{*}=0$. Cross-sections at forward $y^{*}$ (proton or deuteron direction) are significantly smaller than at backward $y^{*}$ (heavy-ion direction). This asymmetry is quantified using the forward-backward production ratio $R_{\textrm{FB}}$, 
\begin{equation}
R_{\textrm{FB}}(\pt, y^{*}) \equiv \frac{d^{2}\sigma(\pt, y^{*}>0)/d\pt dy^{*}}{d^{2}\sigma(\pt, y^{*}<0 )/d\pt dy^{*}}.
\label{eq:RFB}
\end{equation}
This observable has the advantage that it does not rely on knowledge of the $\jpsi$ production cross-section in $pp$ collisions, and that experimental and theoretical uncertainties partially cancel in the ratio. The LHCb Collaboration has recently measured $R_{\textrm{FB}}$ in the range ${2.5<|y^{*}|<4.0}$, ${0<\pt<14\GeV}$~\cite{Aaij:2013zxa}. Results for prompt $\jpsi$ production show a strong $\pt$ dependence with $R_{\textrm{FB}}$ values significantly below unity. In contrast, the $R_{\textrm{FB}}$ for nonprompt $\jpsi$ is consistent with unity and with no $\pt$ dependence. These results are consistent with the measurements presented by the ALICE Collaboration~\cite{Abelev:2013yxa} that do not separate prompt and nonprompt $\jpsi$ production.

This paper presents measurements of differential cross-sections for prompt and nonprompt $\jpsi$ production in {\em p}+Pb collisions at $\sqrt{s_{NN}}$ = 5.02$\TeV$. The kinematic region measured spans the range $8<\pt<30\GeV$ and $-2.87<y^{*}<1.94$. The $\jpsi$ mesons are reconstructed using the dimuon decay mode. Nonprompt $\jpsi$ are separated from prompt $\jpsi$ by measuring displaced decay vertices. $R_{\textrm{FB}}$ measured in the range $|y^{*}|<1.94$ is presented as a function of $\jpsi$ $\pt$ and $y^{*}$. 

ATLAS has previously published measurements of differential cross-sections for $\jpsi$ production in $pp$ collisions at $\sqrt{s}= 7\TeV$~\cite{Aad:2011sp}. This paper uses the methods described in that publication.  

\section{The ATLAS detector}
The ATLAS detector~\cite{Aad:2008zzm} is designed to measure the properties of a wide range of physics processes in {\em pp}, {\em p}+Pb, and Pb+Pb interactions. It has cylindrical geometry and nearly $\rm 4 \pi$ solid-angle coverage.

The inner detector (ID) covers the pseudorapidity\footnote{ATLAS uses a right-handed coordinate system with its origin at the nominal interaction point (IP) in the center of the detector and the $z$-axis along the beam pipe. The $x$-axis points from the IP to the center of the LHC ring, and the $y$-axis points upward. Cylindrical coordinates (r,$\rm \phi$) are used in the transverse plane, $\rm \phi$ being the azimuthal angle around the beam pipe, measured from the $x$-axis. The pseudorapidity is defined in terms of the polar angle $\theta$ as $\eta = -\ln [\tan(\theta/2)]$. Transverse momentum and energy are defined in the $x$-$y$ plane, as $\pt = p \cdot \sin\theta$ and $E_{\mathrm{T}} = E \cdot \sin\theta$.} range $\rm |\eta| < 2.5$ and consists of multiple layers of silicon pixel and microstrip detectors as well as a straw-tube transition radiation tracker (TRT) that covers the range $\rm |\eta| < 2$. The ID is surrounded by a superconducting solenoid that provides a 2 T axial magnetic field.

The calorimeter system surrounds the ID and the solenoid and covers the pseudorapidity range $\rm |\eta| < 4.9$. It provides an excellent containment of electromagnetic and hadronic showers. 

The muon spectrometer (MS) surrounds the calorimeters and consists of multiple layers of trigger and tracking chambers immersed in an azimuthal magnetic field produced by three air-core superconducting magnet systems with average field integrals between 2 and 6 Tm. Drift tubes and cathode strip chambers provide an independent, precise measurement of muon track momentum for $\rm |\eta| < 2.7$. Resistive plate chambers and thin gap chambers provide fast triggering in the range $\rm |\eta| < 2.4$.

The minimum-bias trigger scintillators (MBTS) consist of two sets of sixteen scintillator counters installed on the front face of the endcap calorimeter cryostats. They are used to trigger on minimum-bias events.

A three-level trigger system is employed. The Level-1 trigger is implemented in hardware, using a subset of detector information to reduce the event rate to the design value of 75 kHz. This is followed by two software-based trigger levels, called Level-2 and the Event Filter. For this analysis, the Level-1 trigger and the Event Filter are actively used, while the Level-2 trigger simply passed the events through. 

\section{Data and Monte Carlo samples}
\label{sec:data}
The measurements presented in this paper are performed with a data sample corresponding to an integrated luminosity of 28.1 $\mathrm{nb}^{-1}$ collected in the 2013 LHC {\em p}+Pb run at a center-of-mass energy per nucleon-nucleon pair of $\sqrt{s_{NN}}$ = 5.02$\TeV$. The beams had different energies ($E_{p}$ = 4$\TeV$, $E_{\mathrm{Pb}} = 1.58$ $A$TeV) due to the LHC two-in-one magnet system. Due to this energy difference, the center-of-mass of the proton-nucleon collision system had a longitudinal rapidity shift relative to the ATLAS rest frame of $\Delta y = 0.47$ in the direction of the proton beam. The data was collected in two periods with different beam directions. The typical value for the mean number of interactions per bunch crossing, $\langle \mu\rangle$, was of the order of 0.1.

The luminosity is calibrated by using dedicated beam-separation scans, also known as van der Meer scans~\cite{vanderMeer:1968zz}. Separate calibrations were performed for each period. A systematic uncertainty of $2.7\%$ on the luminosity is evaluated using techniques similar to those described in Ref.~\cite{Aad:2013ucp}. The first period provided approximately $55\%$ of the integrated luminosity, and the proton beam circulated from positive to negative $\eta$; the beam directions were reversed in the second period. 

Monte Carlo (MC) simulations are used to study trigger and reconstruction efficiencies, and kinematic acceptance corrections. \textsc{Pythia8}~\cite{Sjostrand:2007gs} is used to generate $pp$ hard-scattering events in which $\jpsi$ mesons are produced unpolarized either via prompt production or through the decay of $b$-hadrons and subsequently decayed into muon pairs. The detector response is modeled using a {\textsc{Geant4}-based} simulation of the ATLAS detector~\cite{Agostinelli:2002hh,Aad:2010ah}. The events are reconstructed using the same algorithms that were applied to the data. Two separate MC data sets were generated, matching the two different sets of beam directions present in data. The momentum four-vectors of the generated particles are longitudinally boosted by a rapidity $\Delta y =\pm$0.47 to match the corresponding center-of-mass rapidity shift. An additional sample with a large number of simulated $\jpsi\to\mu^{+}\mu^{-}$ events produced unpolarized is used to determine the fiducial acceptance. 

\section{Event and candidate selection}
Proton-lead collisions used in this analysis are selected with a dimuon trigger. The Level-1 trigger requires a single muon with a $\pt$ threshold determined by the largest possible geometrical coincidence between hits from different muon trigger detector layers. The Event Filter performs muon reconstruction using the information from all the detector elements, independently of the Level-1 measurement. Then, it requires at least two muons, each with $\pt >$ 2$\GeV$.

Charged-particle tracks are reconstructed in the ID using an algorithm optimized for minimum-bias measurements in $pp$ collisions~\cite{Aad:2010ac}. The muon candidates are formed from reconstructed ID tracks matched to tracks reconstructed in the MS. The muon ID tracks are required to have at least one pixel detector hit and at least five hits in the microstrip detectors. A successful track extrapolation to the TRT is required for $|\eta|<2$.
Each muon is required to have $|\eta|<2.4$ and $\pt>4\GeV$ and to match the track of a muon reconstructed by the Event Filter;
this matching is performed by requiring the angular separation between the reconstructed and trigger muons to be $\sqrt{(\Delta\eta)^{2} + (\Delta\phi)^{2}} \leq 0.02$. Each muon pair is fit to a common vertex, and a loose requirement on the $\chi^{2}$ of the fit is imposed; MC simulations show that this requirement is fully efficient for $\jpsi\to \mu^{+}\mu^{-}$ decays. The dimuon invariant mass is calculated from the track parameters obtained from the common vertex fit. 

The nonprompt $\jpsi$ are distinguished from prompt $\jpsi$ candidates that are produced either in the primary interaction or in the decay of heavier charmonium states using the ``pseudoproper time,'' $\tau$, defined as
\begin{equation}
\tau = L_{xy} \frac{m_{\mu\mu}}{\pt} ,
\end{equation}
where $m_{\mu\mu}$ is the invariant mass of the dimuon, $\pt$ is its transverse momentum, and $L_{xy}$ is the signed transverse distance between the primary interaction vertex and the $\jpsi\to\mu^{+}\mu^{-}$ vertex. The primary interaction vertex is defined as the vertex with the highest summed $\pt^{2}$ of associated tracks, with the two muon tracks excluded.
The number of events with more than one hard-scattering is not significant due to the beam conditions described in Sec.~\ref{sec:data}; therefore the probability to assign an incorrect primary vertex is neglected.

Dimuons with an invariant mass in the interval ${2.5<m_{\mu\mu}<3.5\GeV}$ are considered $\jpsi$ candidates. This choice excludes the $ \psi(2S)$ region while retaining the regions adjacent to the $\jpsi$ peak to constrain the background shape. 
Possible sources of background include: oppositely charged muons coming from heavy-flavor decays, pairs coming from the Drell-Yan process, and random combinations of muons and hadrons misidentified as muons.

\section{$\jpsi$ signal extraction }
Corrections are applied to the data to account for trigger and reconstruction efficiencies and kinematic acceptance. Each $\jpsi$ candidate is assigned a weight, $w$, defined as
\begin{equation}
w^{-1} = \mathcal{A} \cdot \epsilon_{\mathrm{reco}}\cdot \epsilon_{\mathrm{trigger}} ,
\label{eq:weight}
\end{equation}
where $\mathcal{A}$ is the kinematic acceptance, $\epsilon_{\mathrm{reco}}$ is the dimuon reconstruction efficiency, and $\epsilon_{\mathrm{trigger}}$ is the trigger efficiency. The use of per-candidate weights avoids potential biases that may result from the variation of these quantities over the kinematic intervals used in the analysis. 

The kinematic acceptance is defined as the fraction of $\jpsi\to\mu^{+}\mu^{-}$ decays for which both muons have $\pt>4\GeV$ and $|\eta|<2.4$. The dimuon reconstruction efficiency is defined as the probability that a $\jpsi$ satisfying the acceptance criteria passes the offline reconstruction requirements. The trigger efficiency is defined as the probability for events containing reconstructed $\jpsi$ candidates to pass the trigger selections.   

The kinematic acceptance is derived in fine intervals of $\jpsi$ $\pt$ and $y$ using a generator-level MC simulation of unpolarized $\jpsi\to\mu^{+}\mu^{-}$ decays. 

The dimuon reconstruction efficiency is assumed to be given by the product of two single-muon reconstruction efficiencies $\epsilon_{\mathrm{reco}}^{\mu}$, 
\begin{equation}
 \epsilon_{\mathrm{reco}} =  \epsilon_{\mathrm{reco}}^{\mu} ( p^{\mu}_{\textrm{T1}}, q_{1}^{\mu} \cdot \eta^{\mu}_{1} ) \cdot  \epsilon_{\mathrm{reco}}^{\mu} ( p^{\mu}_{\textrm{T2}}, q_{2}^{\mu} \cdot \eta^{\mu}_{2} ) ,
\end{equation}
where $\pt^{\mu}$, $q^{\mu}$, and $\eta^{\mu}$ are transverse momentum, charge, and pseudorapidity of the muons. The $\epsilon_{\mathrm{reco}}^{\mu}$ is derived from $pp$ data using $\jpsi\to\mu^{+}\mu^{-}$ decays, as described in Ref.~\cite{Aad:2014rra}.

The Level-1 trigger efficiency $\epsilon_{\mathrm{L1}}$ is defined as the probability that an event passing the reconstruction requirements is selected by the Level-1 trigger. The Event Filter efficiency $\epsilon_{\mathrm{EF}}$ is defined as the probability that events selected by the Level-1 trigger are selected by the Event Filter. Because the Event Filter performs muon reconstruction independently of the Level-1 trigger, the trigger efficiency is calculated as
\begin{equation}
\epsilon_{\mathrm{trigger}} =  \epsilon_{\mathrm{L1}} \cdot \epsilon_{\mathrm{EF}} .
\label{eq5}
\end{equation}
The efficiency $\epsilon_{\mathrm{L1}}$ is expressed in terms of the single-muon Level-1 efficiency $\epsilon_{\mathrm{L1}}^{\mu}$. The Level-1 trigger required at least one muon in the event, thus
\begin{equation}
\epsilon_{\mathrm{L1}} = 1-[1-\epsilon_{\mathrm{L1}}^{\mu}( p^{\mu}_{\textrm{T1}}, q_{1}^{\mu} \cdot \eta^{\mu}_{1} )]\cdot [1-\epsilon_{\mathrm{L1}}^{\mu}( p^{\mu}_{\textrm{T2}}, q_{2}^{\mu} \cdot \eta^{\mu}_{2})].
\label{eq6}
\end{equation}
The efficiency $\epsilon_{\mathrm{L1}}^{\mu}$ is derived from data using reconstructed muons in events selected with a minimum-bias trigger that required a signal in at least one MBTS counter on each set. It is defined as the ratio of the number of reconstructed muons that passed the trigger requirement to the number of reconstructed muons in each $\pt^{\mu}$ and $q^{\mu} \cdot \eta^{\mu}$ interval.   

The efficiency $\epsilon_{\mathrm{EF}}$ is expressed in terms of the single-muon Event Filter efficiency $\epsilon_{\mathrm{EF}}^{\mu}$. The Event Filter selected events with two muons, thus  
\begin{equation}
\epsilon_{\mathrm{EF}} = \epsilon_{\mathrm{EF}}^{\mu}( p^{\mu}_{\textrm{T1}}, q_{1}^{\mu} \cdot \eta^{\mu}_{1} )\cdot  \epsilon_{\mathrm{EF}}^{\mu}( p^{\mu}_{\textrm{T2}}, q_{2}^{\mu} \cdot \eta^{\mu}_{2} ).
\label{eq7}
\end{equation}

The efficiency $\epsilon_{\mathrm{EF}}^{\mu}$  is determined from MC simulation and checked with data; in both cases the ``tag and probe'' method is used. In this method, events selected with single-muon triggers with various thresholds starting from $\pt^{\mu}> 4$ $\GeV$ are used to select muon pairs by requiring a well-reconstructed muon, the ``tag,''  and another muon, the ``probe,'' that form a pair consistent with originating from a $\jpsi$ decay. The tag is required to be consistent with the particle that triggered the event and to pass the Level-1 requirement. The probes provide a sample that can be used to measure the trigger efficiency in an unbiased way. The Event Filter efficiency $\epsilon_{\mathrm{EF}}^{\mu}$ is evaluated as the ratio of the number of $\jpsi$ (determined by fitting the $m_{\mu\mu}$ distributions) with probes that pass the Event Filter requirements, to the total number of selected $\jpsi$. Results from MC simulation and data agree within the statistical uncertainty of the data.  

\begin{figure}[htb]
\includegraphics[width=.49\textwidth]{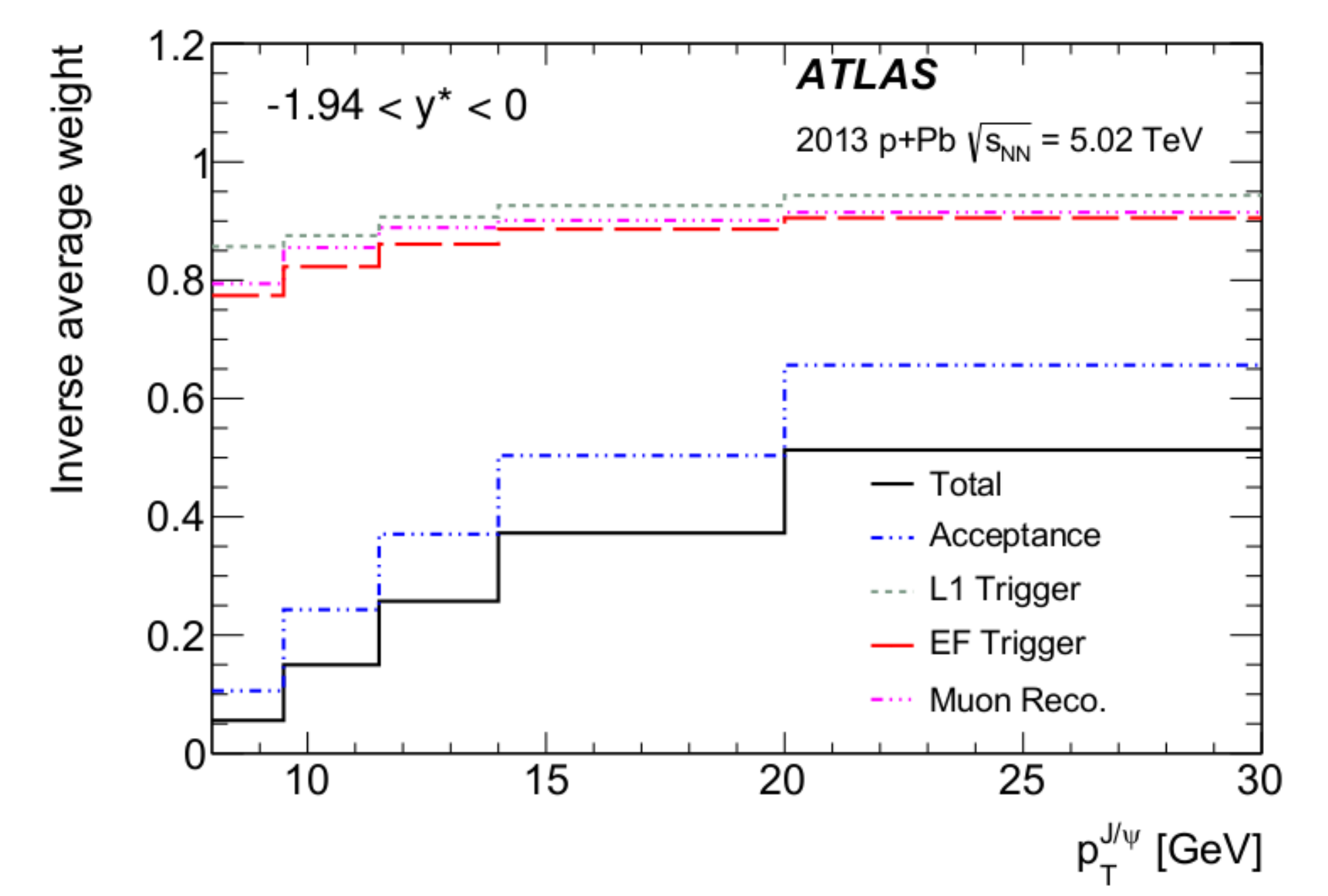}\\
\includegraphics[width=.49\textwidth]{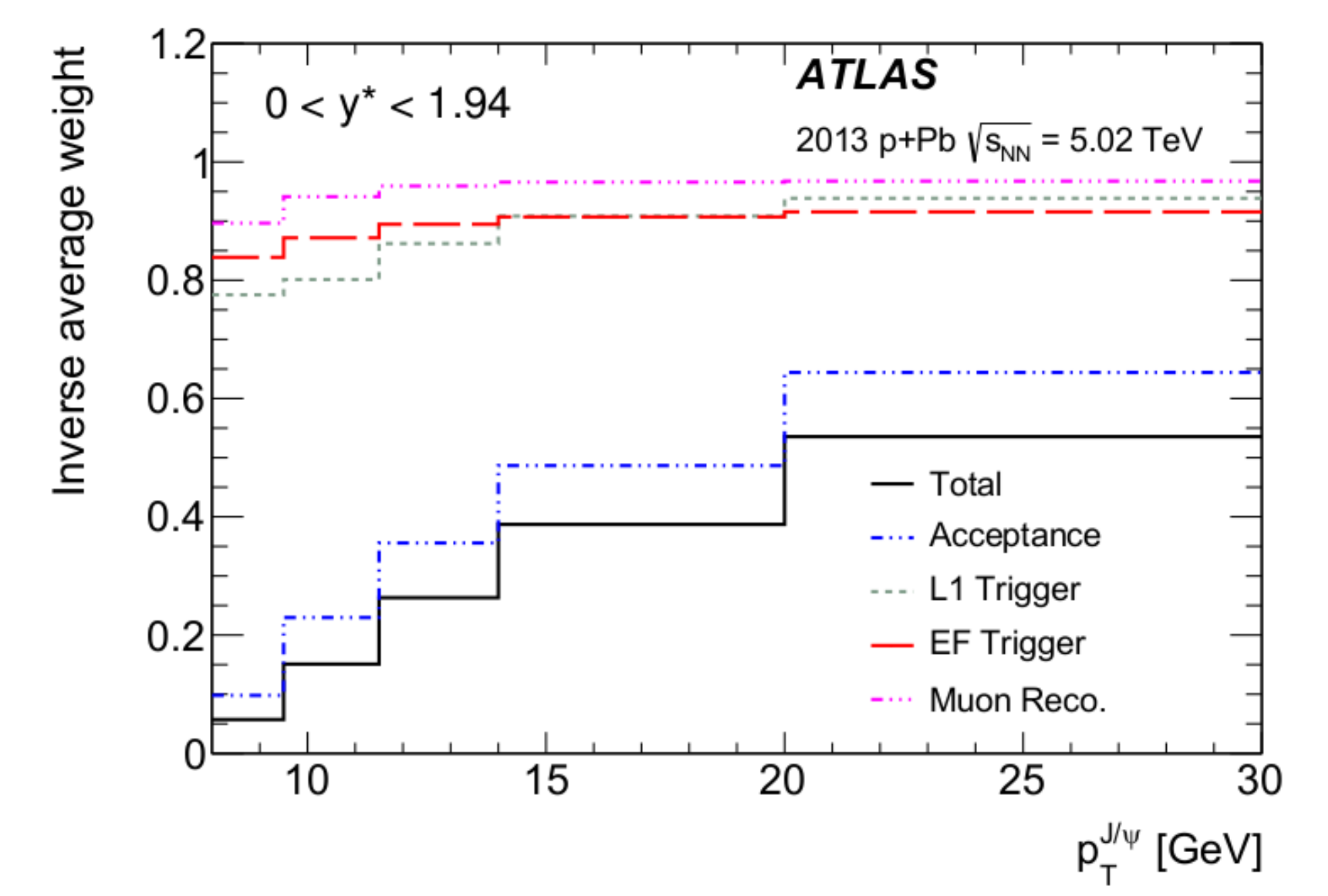}\\
\includegraphics[width=.49\textwidth]{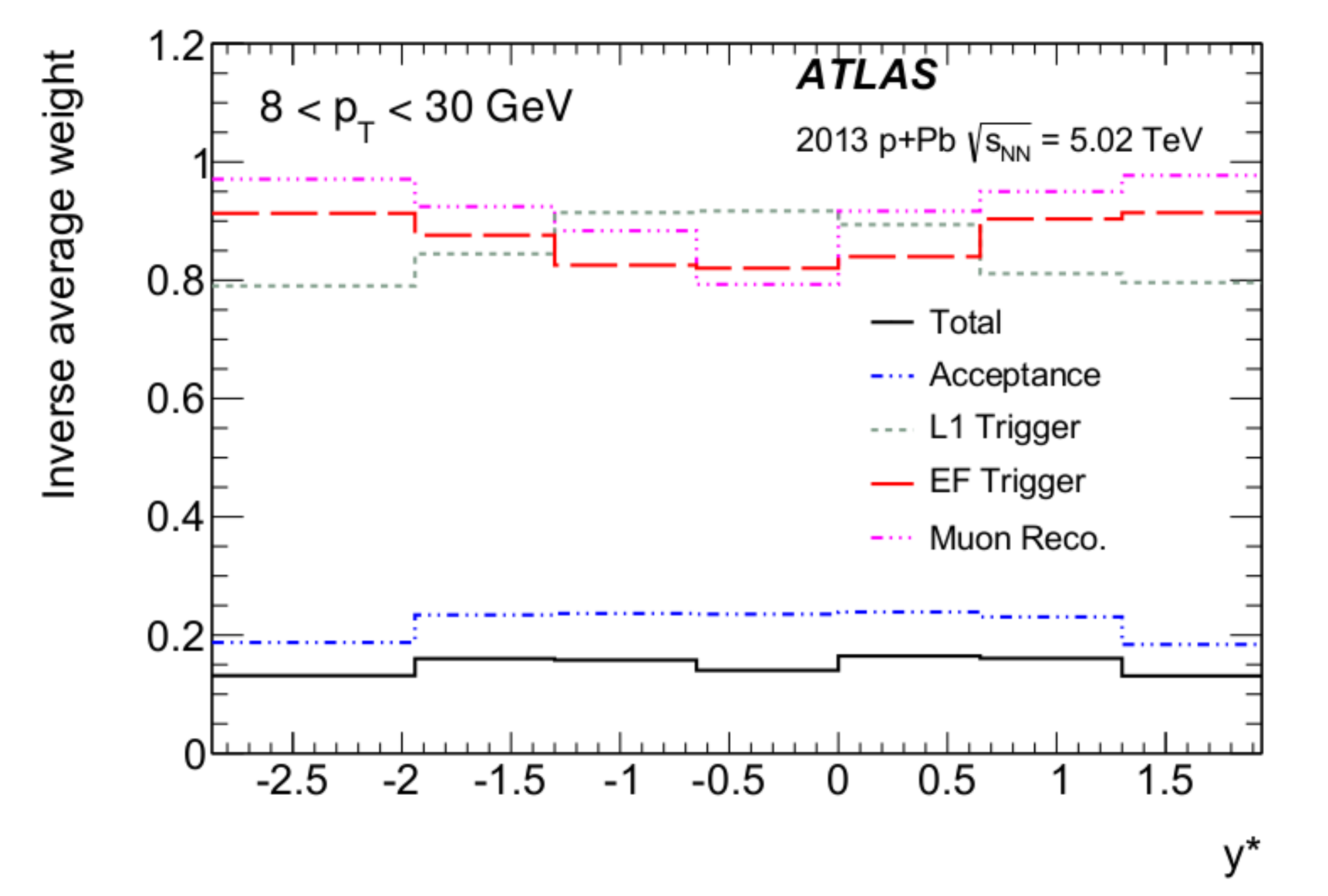}
\caption{The inverse of the average weight for $J/\psi$ candidates as a function of $J/\psi$ transverse momentum and center-of-mass rapidity. The relative contributions from kinematic acceptance, reconstruction, and trigger corrections are also shown. The weights are extracted from a combination of data and MC simulation.}
\label{fig:inverse}
\end{figure}

The data are corrected on a per-candidate basis, using the weights defined in Eq.~\eqref{eq:weight}. To illustrate the impact of the corrections, the average weights over all $J/\psi$ candidates evaluated for the kinematic intervals used in the cross-section measurement are shown in Fig.~\ref{fig:inverse}. The relative contributions from the kinematic acceptance and the trigger and reconstruction efficiencies are shown separately. Due to the center-of-mass boost, the intervals of $y^{*}$ used for the forward-backward asymmetry measurement span intervals in $y$ that are not symmetric around $y=0$. Those intervals are listed in Table~\ref{table:interval}. In both periods the $\jpsi$ candidates with $|y|<0.47$ are in the negative $y^{*}$ interval, whereas those with $1.47<|y|<2.4$ are in the positive $y^{*}$ interval. As a result, the weights obtained for the positive and negative $y^{*}$ intervals are different.

\begin{table}[h]
  \caption{Intervals of rapidity in the ATLAS reference frame for $-1.94< y^{*} <0$ and $0< y^{*} <1.94$ for the two run periods with different beam directions. The center-of-mass shift corresponds to $\Delta y= 0.47$ in the proton-beam direction.}
    \begin{tabular}{
l@{\extracolsep{10pt}}
c@{\extracolsep{20pt}}
c@{\extracolsep{20pt}}
}
\hline
\hline
                     &   $-1.94< y^{*} <0 $ &  $  0< y^{*} <1.94$ \\
First period   &  $-0.47<y<+1.47  $ & $  -2.4<y<-0.47$\\
Second period & $-1.47<y<+0.47  $ & $  +0.47<y<+2.4$ \\
\hline
\hline
\end{tabular}
\label{table:interval}
\end{table}

\begin{figure}[!htb]
\includegraphics[width=.47\textwidth]{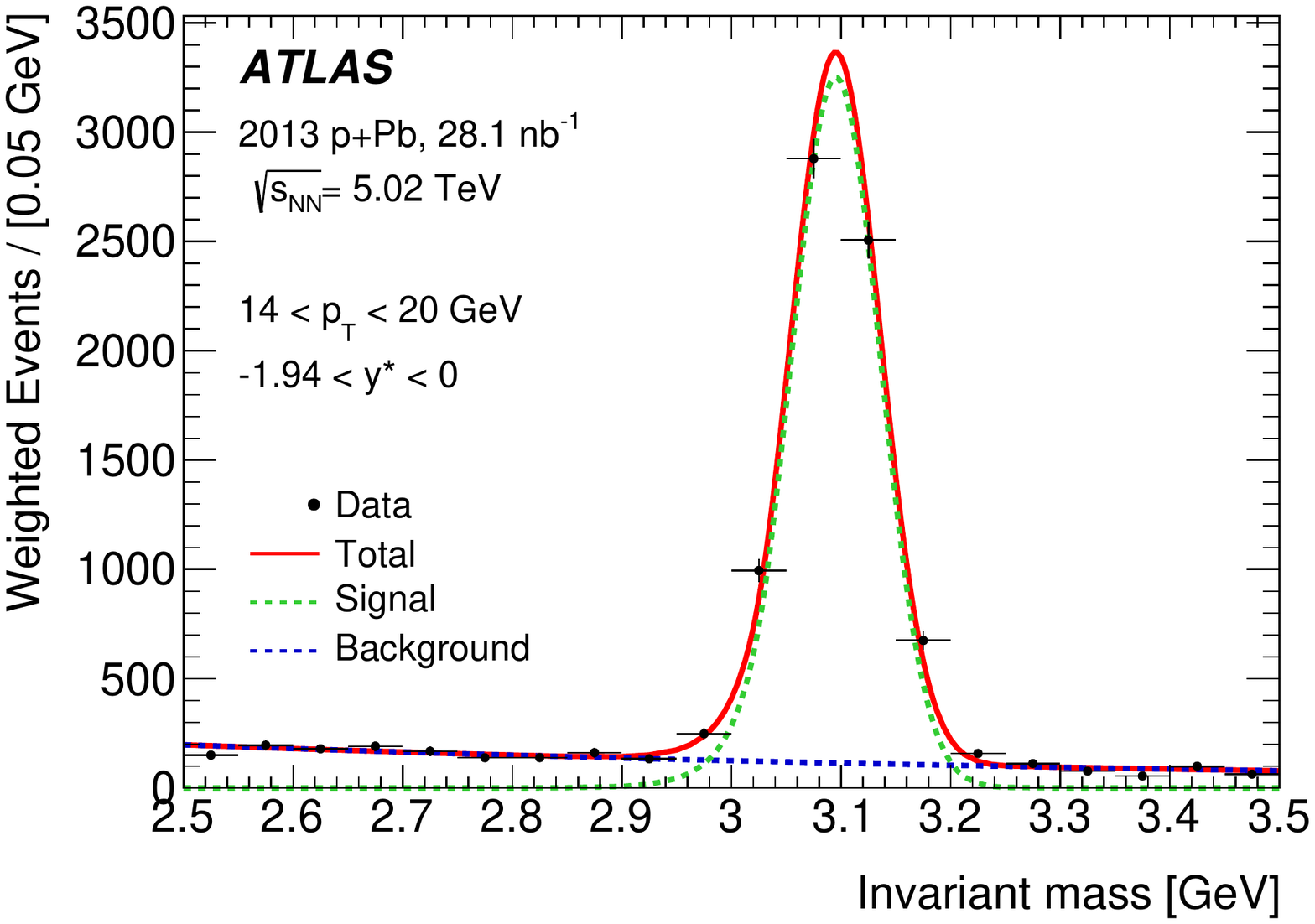}
\includegraphics[width=.47\textwidth]{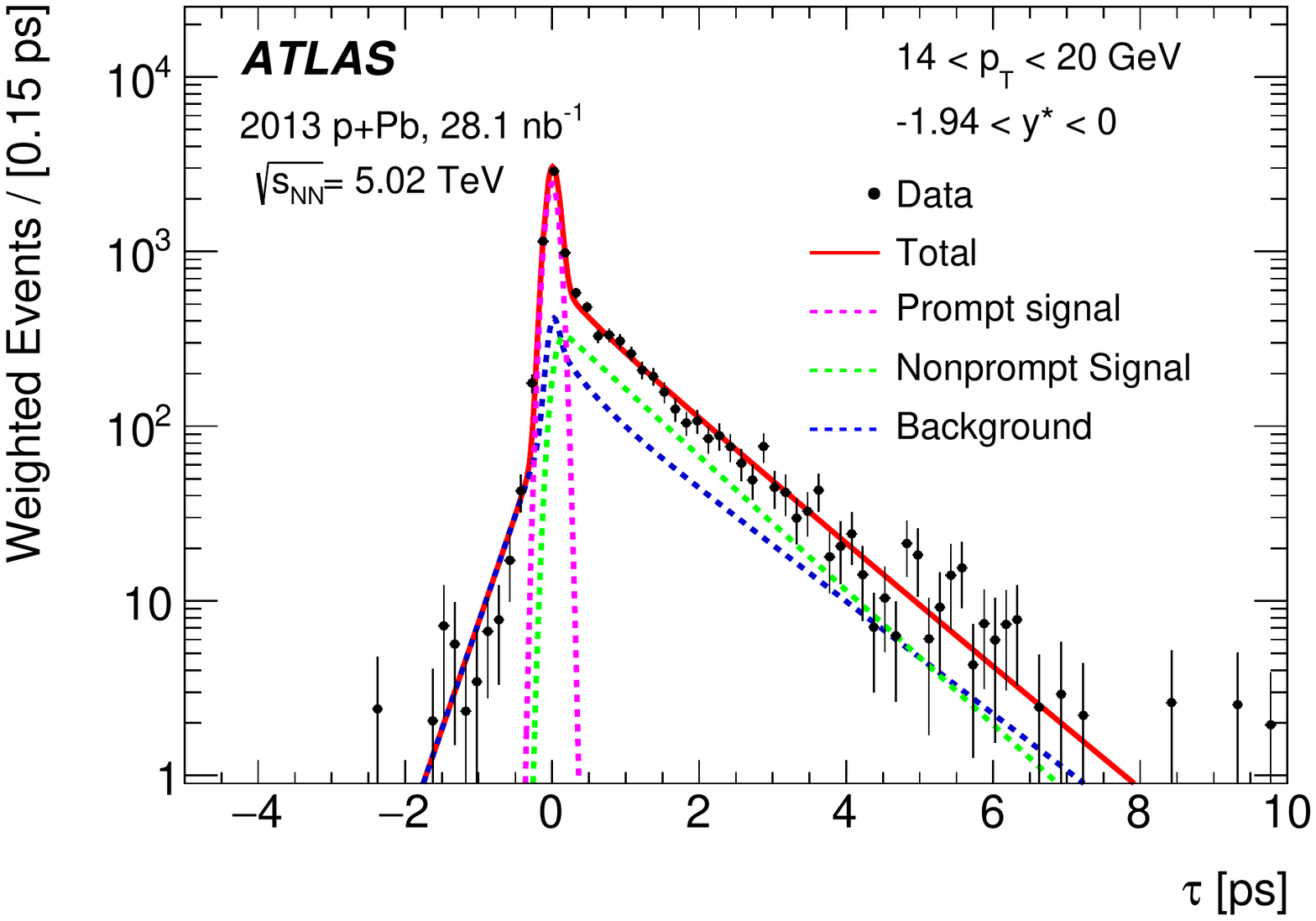}
\caption{Distributions of dimuon invariant mass (upper panel) and pseudoproper time (bottom panel) of weighted $J/\psi$ candidates in a representative interval of $J/\psi$ transverse momentum and center-of-mass rapidity. The projection of the function resulting from a two-dimensional unbinned maximum-likelihood fit is also shown.}
\label{fig:fitexample}
\end{figure}

The number of produced $\jpsi$ mesons and the relative fraction of nonprompt $\jpsi$ with respect to inclusive production, called the ``nonprompt fraction,'' are determined using a two-dimensional extended maximum-likelihood fit~\cite{Barlow1990496} of the ($m_{\mu\mu}, \tau$) spectrum of weighted $\jpsi$ candidates. The fit functions used are similar to those described in previous ATLAS publications~\cite{Aad:2011sp}. The signal $\tau$ distribution is described using a Dirac delta function for prompt $\jpsi$ and an exponential function for nonprompt $\jpsi$; these are convolved with a Gaussian resolution function whose width is a free parameter. The background $\tau$ distribution is described with the sum of a delta function to describe prompt background, an exponential function to describe nonprompt background, and a double-sided exponential function to describe non-Gaussian tails observed at negative $\tau$; these are convolved with a Gaussian resolution function whose width is a free parameter not restricted to be the same as the signal resolution. The $m_{\mu\mu}$ spectrum is described by a ``Crystal Ball'' (CB) function~\cite{Oreglia:1980cs} for the signal and an exponential function for the background. The complete fit model includes 15 free parameters. Fits are performed using MINUIT~\cite{James:1975dr} interfaced with the RooFit~\cite{Verkerke:2003ir} framework. The fit is performed separately in several bins of dimuon $\pt$ and $y^{*}$. Figure~\ref{fig:fitexample} shows $m_{\mu\mu}$ and $\tau$ distributions in the kinematic interval $14<\pt<20\GeV$, $-1.94<y^*<0$, and the corresponding projections of the fit function.  

Several studies with pseudoexperiments and other cross-checks show that the fit procedure provides an unbiased estimation of the extracted parameters and their statistical uncertainties.  
\begin{table*}[t]
  \caption{Summary of statistical and systematic uncertainties on the differential cross-section measurements for prompt and nonprompt $J/\psi$. The values are quoted as relative uncertainties (in $\%$) and refer to the range of uncertainties over the specified $p_{\textrm{T}}$ or $y^{*}$ range.}
    \begin{tabular*}{0.9\textwidth}{
l@{\extracolsep{40pt}}
c@{\extracolsep{40pt}}
c@{\extracolsep{40pt}}
c@{\extracolsep{40pt}}
}
    \hline    
\hline      
 Uncertainty&    $-1.94< y^{*} <0$ &  $0< y^{*} <1.94$ & $8< p_{\mathrm{T}}<30$ GeV\\
  &     $p_{\mathrm{T}}$ range [8,30] GeV & $p_{\mathrm{T}}$ range [8,30] GeV & $y^{*}$ range [$-$2.87,1.94] \\
Statistical  & 2.1--5.9 & 2.3--6.9 &2.6--10  \\
Trigger & 5.3--7.5 & 5.2--7.4 & 5.7--7.0 \\
Muon Reconstruction & 2.6--4.2 & 2.4--3.7 & 2.2--3.6 \\
Fit Model & 3.3--6.1 & 2.4--9.2 &2.9--17 \\
Luminosity & 2.7& 2.7 &2.7\\
\hline
\hline
\end{tabular*}
\label{table:systematics}
\end{table*}

\newpage

\section{Systematic uncertainties}
\label{sec:systematics}
The relevant sources of systematic uncertainty for the measurements presented in this work are trigger and reconstruction efficiency corrections, fit model dependence, and the luminosity calibration. 

The dominant source of systematic uncertainty associated with the Event Filter efficiency is the limited size of the data sample available for the tag-and-probe study. The corresponding systematic uncertainty on the cross-section measurement is estimated by means of pseudoexperiments, randomly varying the weight used for each $\jpsi$ candidate according to the uncertainty in the single-muon efficiency. 

The systematic uncertainty associated with the Level-1 trigger efficiency is estimated by varying the selection criteria for muons and by considering discrepancies with an alternative determination of the efficiency using MC simulation. 

The systematic uncertainties associated with muon reconstruction efficiencies were evaluated in Ref.~\cite{Aad:2014rra} using 2012 $pp$ data. Detector operating conditions and occupancy were similar in the 2012 $pp$ run and the 2013 {\em p}+Pb run; therefore the efficiencies and uncertainties calculated in Ref.~\cite{Aad:2014rra} are used in the present analysis.

The impact of the Level-1 trigger and muon reconstruction systematic uncertainties on the $\jpsi$ cross-section is estimated by varying all of the efficiency corrections up and down by their systematic uncertainties, and re-calculating the mean dimuon reconstruction efficiency over all $\jpsi$ candidates in each kinematic bin. The resulting deviation of the mean dimuon reconstruction efficiency from the central value in each bin is taken as a systematic uncertainty on the $\jpsi$ inclusive cross-section.

A closure test of the overall trigger efficiency corrections is performed by means of MC simulations. The result indicates that the assumption of factorization in Eqs.~\eqref{eq5} to~\eqref{eq7} results in a bias of 2--5$\%$ depending on the kinematic bin. This nonclosure is taken as a systematic uncertainty on the $\jpsi$ inclusive cross-section.

The systematic uncertainty associated with the fit model is estimated by varying the fit functions to gauge the sensitivity of the inclusive number of observed $\jpsi$ and the nonprompt fraction to the function chosen for the fits. The signal $m_{\mu\mu}$ distribution is fit with a CB function that can account for the tail observed in the low mass region. A double-Gaussian distribution with different widths but the same mean can adequately describe the signal in most regions, and this is used as a variation. The $m_{\mu\mu}$ distribution of the background is modeled by an exponential function. A second-order Chebyshev polynomial is used as an alternative. The resolution function used for the modeling of both the signal and background $\tau$ distributions is changed to a double-Gaussian function as an alternative. These variations are performed separately.

The variation in the background shape in the $\tau$ distribution is addressed in the following way: a background-only fit is performed to the $\tau$ distribution in a sideband region defined by dimuons with $m_{\mu\mu}$ in the interval of 2.5--2.8$\GeV$ or 3.2--3.5$\GeV$. The background shape parameters are fixed and then the fit is performed in the 2.5--3.5$\GeV$ mass region. 

The systematic uncertainty associated with each fit variation is taken as the deviation from the central value. The total systematic uncertainty of the fit model is taken as the sum in quadrature of the effects of using the alternative fit functions and the fit constrained by the sideband region. It is dominated by the uncertainty associated to the modeling of the $\tau$ distribution.

The luminosity systematic uncertainty of $2.7\%$ is propagated to the differential cross-section measurements presented. It is not considered in the measurement of the nonprompt fraction or the forward-backward ratio as both of these observables are independent of the luminosity.

The kinematic acceptance correction has a potential theoretical uncertainty that depends on the spin-alignment of the $\jpsi$ decay. Previous measurements in {\em pp} collisions~\cite{Abelev:2011md,Aaij:2013nlm,Chatrchyan:2013cla} suggest that the degree of polarization is small at LHC energies. Based on the assumption that the nuclear medium does not modify the average polarization of produced $\jpsi$, no systematic uncertainty due to spin-alignment is included.  The modification to quoted production rates under various benchmark spin-alignments assumptions are presented in in Appendix~\ref{sec:AppendixA}.

The kinematic acceptance correction is obtained using a large sample of MC simulated events that allows the kinematic variables to be binned finely. Therefore, the impact of mismodeling of the underlying kinematic distributions in the MC simulation, as reported in previous ATLAS publications~\cite{Aad:2011sp}, is negligible.

The total systematic uncertainty on the $\jpsi$ inclusive differential cross-section amounts to 6--9$\%$, with no strong $y^{*}$ or $\pt$ dependence, and is dominated by trigger efficiency systematic uncertainties. The systematic uncertainty in the nonprompt fraction, estimated from fit model variations, amounts to 2--17$\%$, with the largest values at large $|y^{*}|$ and low $\pt$. 

The systematic uncertainties on the cross-section for prompt and nonprompt $\jpsi$ are obtained from the systematic uncertainties of the inclusive cross-section and the nonprompt fraction, assuming them to be uncorrelated. The corresponding statistical uncertainties are obtained by considering the covariance between the fit parameters. A  summary of the statistical and systematic uncertainties of the differential cross-section measurements for prompt and nonprompt $\jpsi$ {are shown in Table~\ref{table:systematics}}.

\section{Results and discussion}
\label{sec:results}
\subsection{Cross-sections and nonprompt fraction}
\begin{figure}[htp]
\includegraphics[width=.49\textwidth]{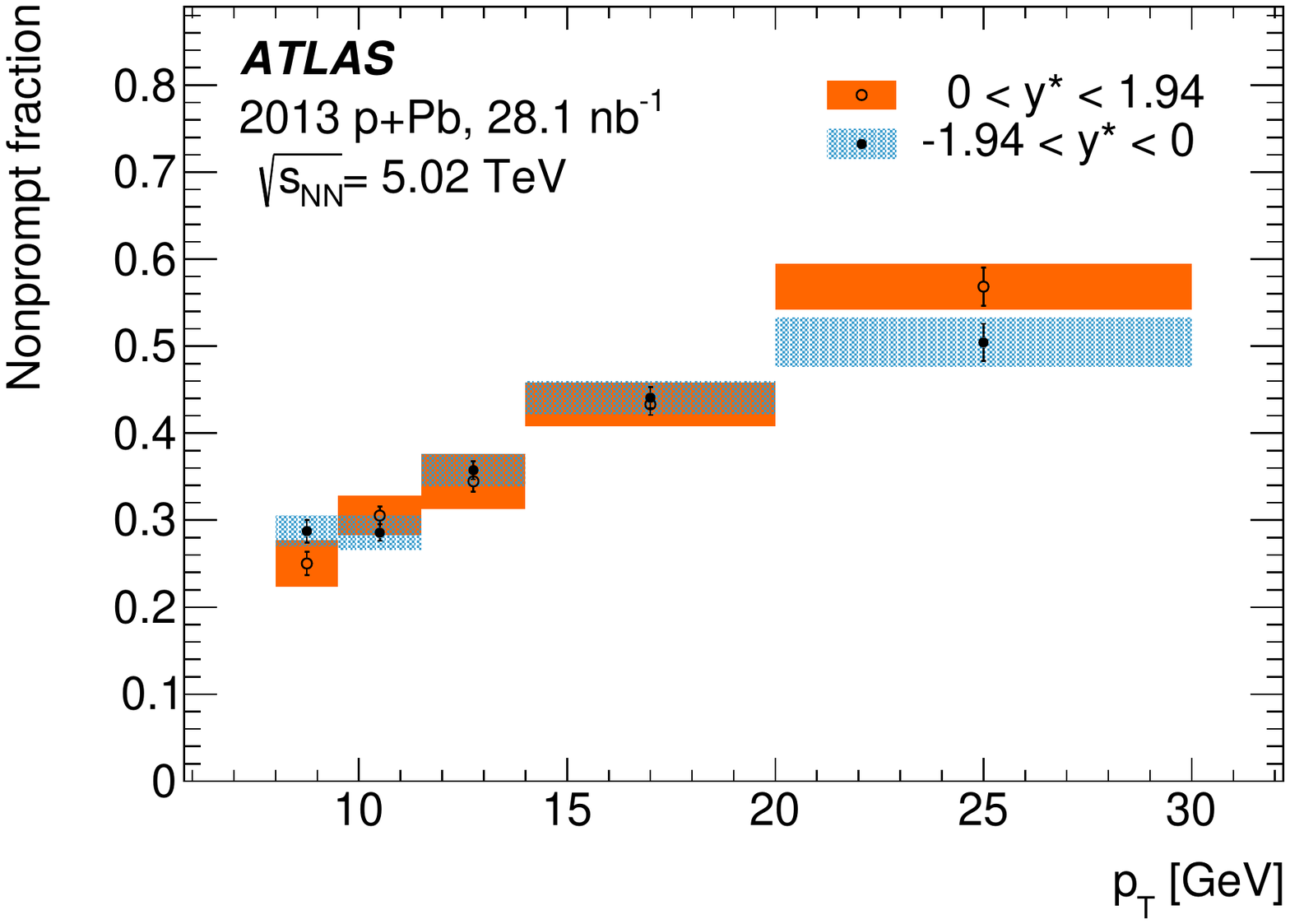}
\includegraphics[width=.49\textwidth]{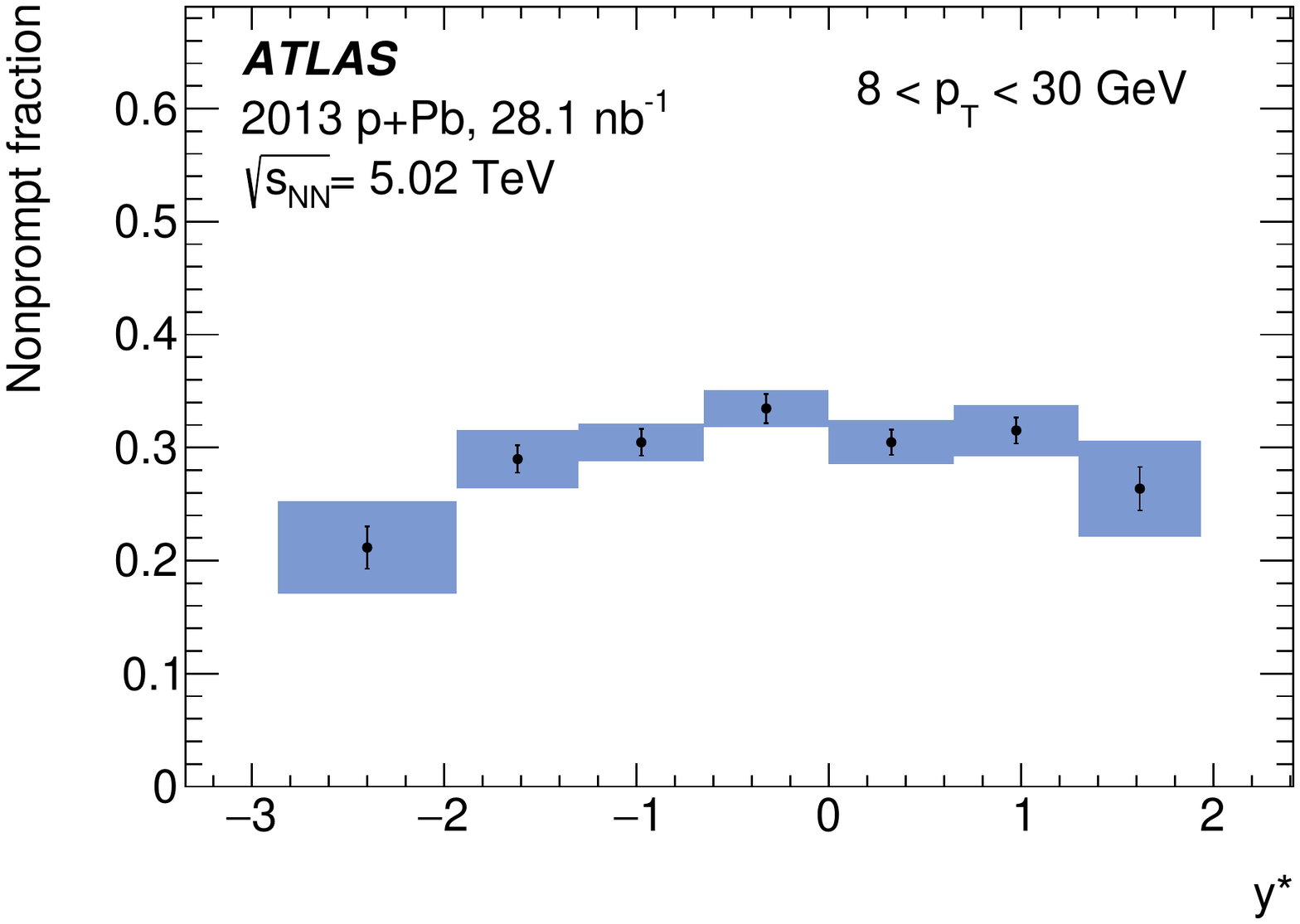}
\caption{Nonprompt fraction as a function of $J/\psi$ transverse momentum $p_{\mathrm{T}}$ (upper panel) and center-of-mass rapidity $y^{*}$ (bottom panel). Positive $y^{*}$ is defined in the proton beam direction. The error bars show the statistical uncertainty, and the shaded boxes show the sum in quadrature of statistical and systematic uncertainties.}
\label{fig:Xsection_Bfraction}
\end{figure}

The measured nonprompt fraction in the backward $(-1.94 < y^{*} < 0)$ and forward $(0 < y^{*} < 1.94)$ regions is shown as a function of $\jpsi$ transverse momentum in the upper panel of Fig.~\ref{fig:Xsection_Bfraction}. 

A strong $\pt$ dependence of the nonprompt fraction is observed, reaching values above 50$\%$ at the highest measured $\pt$. There is no significant difference between the forward and backward $y^{*}$ measurements. The measured nonprompt fraction integrated over the transverse momentum range $8< \pt <30\GeV$ is shown as a function of $y^{*}$ in the bottom panel of Fig.~\ref{fig:Xsection_Bfraction}. No significant $y^{*}$ dependence is observed. Previous measurements~\cite{Aad:2011sp,Khachatryan:2010yr} with $pp$ collisions in a similar kinematic region show similar trends.

The differential cross-sections are defined as
\begin{equation}
\frac{d^{2}\sigma}{d\pt dy^{*}} \times \textrm{BR}( \jpsi \to \mu^{+}\mu^{-} ) = \frac{N_{\mathrm{corr}}^{\jpsi}}{\mathcal{L} \times \Delta \pt \times \Delta y^{*}}
\label{eq:Xsection}
\end{equation}
where $\textrm{BR}( \jpsi\to\mu^{+}\mu^{-} )$ is the branching ratio of the dimuon channel, $N_{\mathrm{corr}}^{\jpsi}$ is the number of observed $\jpsi$ obtained from the fit to the weighted data, $\mathcal{L}$ is the integrated luminosity of the sample, and $\Delta \pt$ and $\Delta y^{*}$ are the transverse momentum and center-of-mass rapidity bin widths. 

\begin{figure}[htb]
\centering 
\includegraphics[width=.49\textwidth]{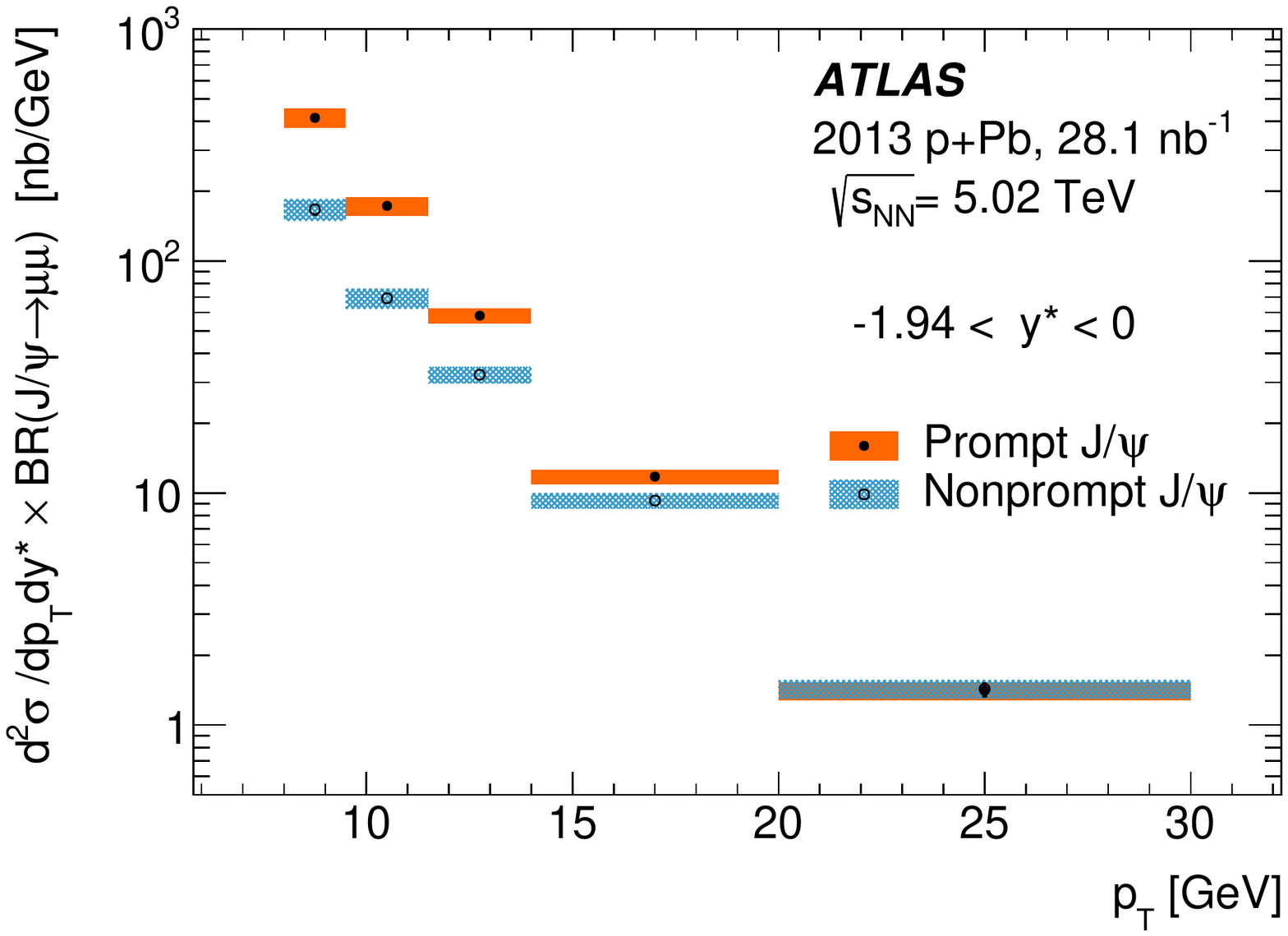}
\includegraphics[width=.49\textwidth]{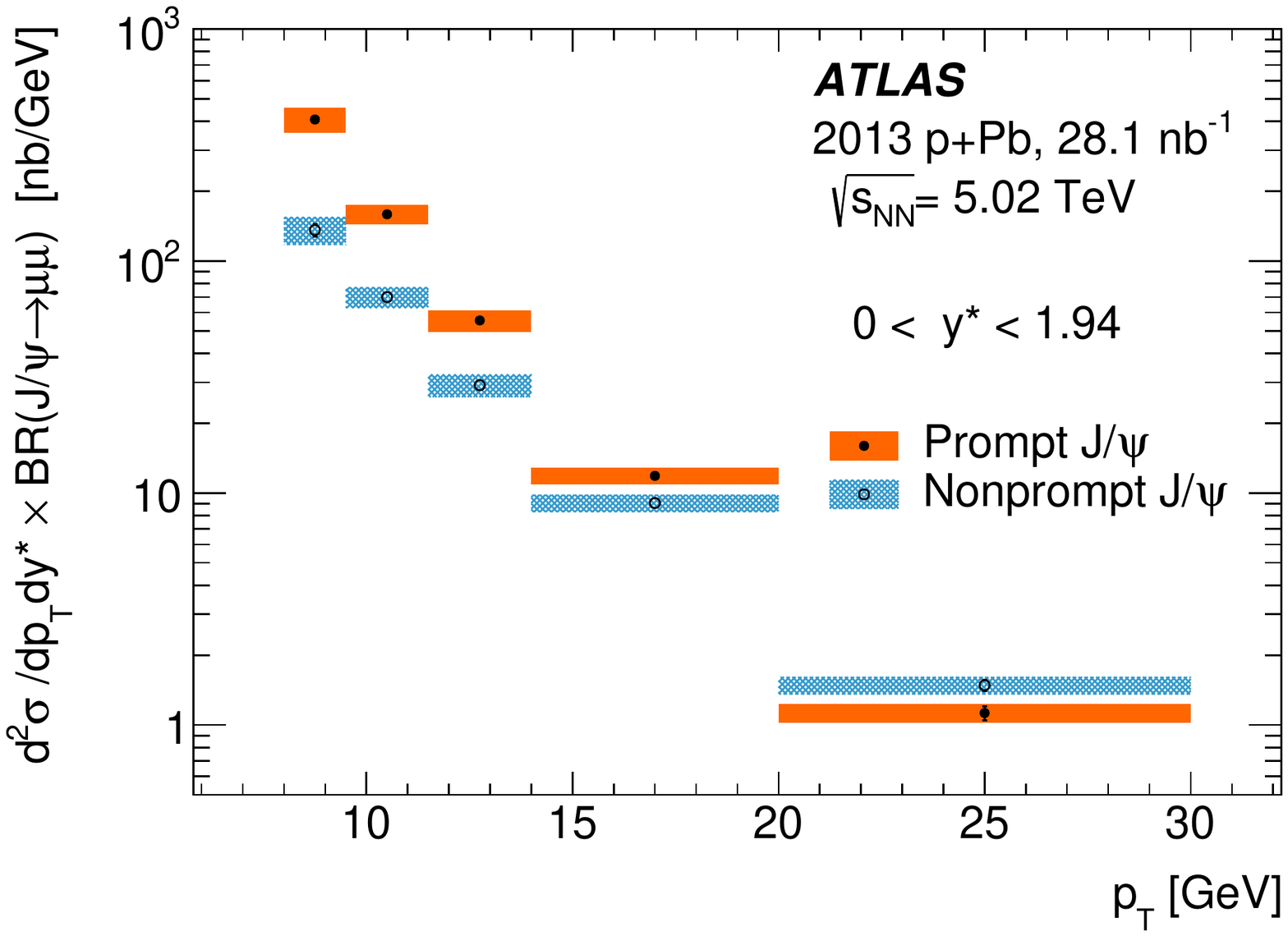}
\caption{Double differential cross-section for prompt and nonprompt $J/\psi$ production as a function of $J/\psi$ transverse momentum, $p_{\mathrm{T}}$. The upper panel shows results in backward $y^{*}$ (lead beam direction), and bottom panel in forward $y^{*}$ (proton beam direction). The error bars show the statistical uncertainty, and the shaded boxes show the sum in quadrature of statistical and systematic uncertainties.}
\label{fig:Xsection_pt}
\end{figure}

\begin{figure}[htp]
\includegraphics[width=.49\textwidth]{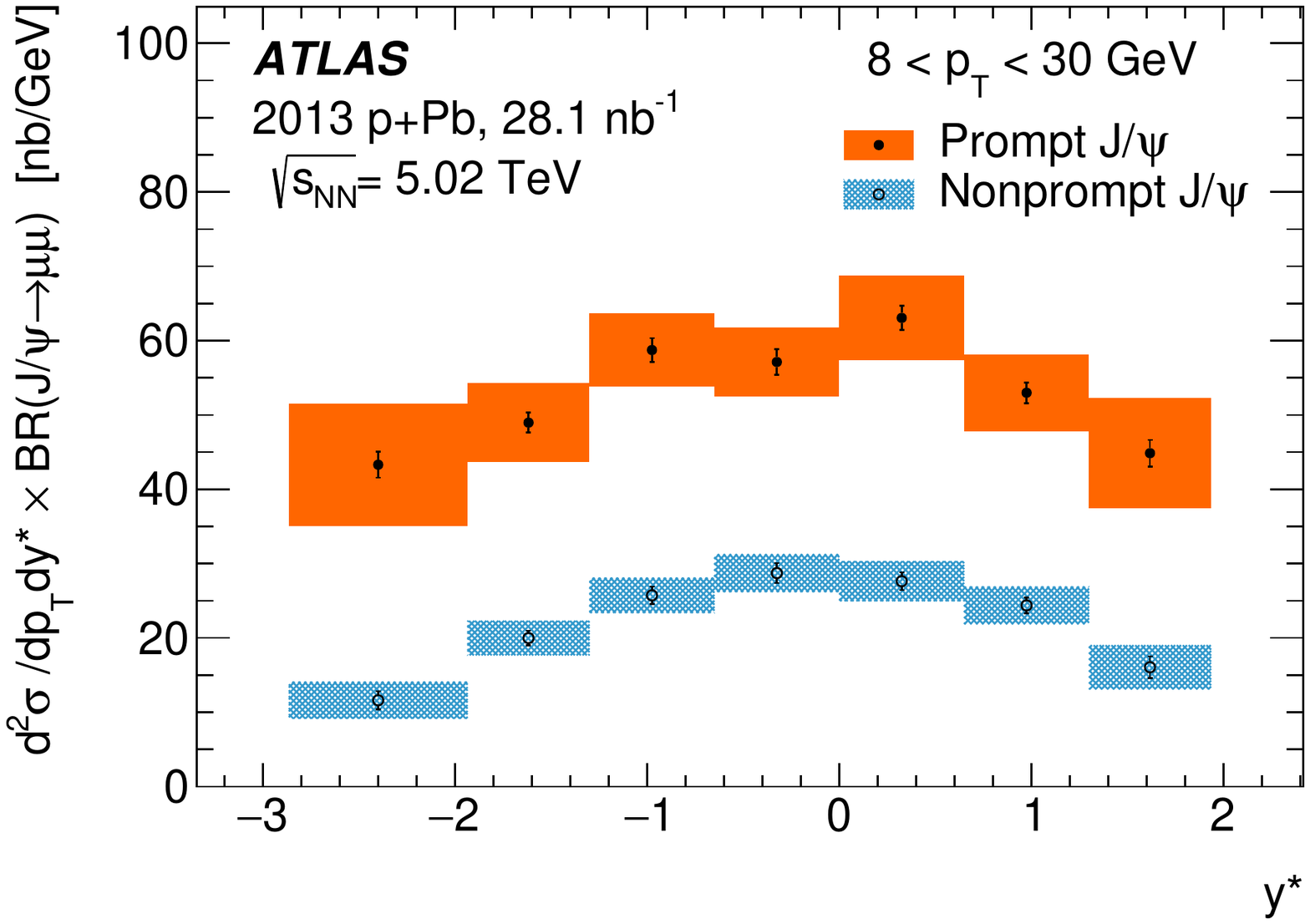}
\caption{Double differential cross-section for prompt and nonprompt $J/\psi$ as a function of $J/\psi$ center-of-mass rapidity $y^{*}$. Positive $y^{*}$ is defined in the proton beam direction. The error bars show the statistical uncertainty, and the shaded boxes show the sum in quadrature of statistical and systematic uncertainties.}
\label{fig:Xsection_rapidity}
\end{figure}

The cross-sections for prompt and nonprompt $\jpsi$ are derived from the inclusive production cross-section and the nonprompt fraction. Differential cross-sections for prompt and nonprompt $\jpsi$ production are shown in Fig.~\ref{fig:Xsection_pt} as a function of $\pt$ in the backward and forward $y^{*}$ regions, and in Fig.~\ref{fig:Xsection_rapidity} as a function of $y^{*}$. The statistical uncertainties are negligible relative to the systematic uncertainties except at high $\pt$. The rapidly falling spectrum and the different slopes for the two production modes are similar to previous measurements~\cite{Aad:2011sp,Khachatryan:2010yr}. No significant asymmetry is observed as a function of $y^{*}$, and the $\pt$ dependence at forward and backward $y^{*}$ is found to be compatible. This is quantified by the ratio $R_{\textrm{FB}}$, as discussed in the following section.  

\subsection{Forward-backward ratio}
\label{sec:RFB}
The asymmetry of $\jpsi$ production between the proton beam direction and lead beam direction is quantified with the forward-backward ratio $R_{\textrm{FB}}$, defined in Eq.~\eqref{eq:RFB}. It is calculated from the cross-section measurements presented in Fig.~\ref{fig:Xsection_pt} and Fig.~\ref{fig:Xsection_rapidity}, and is thus presented integrated over $|y^{*}|<1.94$ as a function of $\pt$, and also integrated over $8<\pt<30\GeV$ as a function of $|y^{*}|$. This ratio is sensitive to a possible rapidity dependence of cold-medium effects in $\jpsi$ production. 

Systematic uncertainties in the forward and backward $y^{*}$ regions partially cancel out in $R_{\textrm{FB}}$, when integrated over $|y^{*}|<1.94$, because $\jpsi$ candidates with exactly the same $y$ fall in either forward or backward $y^{*}$ depending on the beam directions of the data-taking period. As shown in Table~\ref{table:interval}, $\jpsi$ candidates with $0.47<y<1.47$ fall in the backward $y^{*}$ in the first period but in forward $y^{*}$ in the second period. Similarly, $\jpsi$ candidates with $-1.47<y<-0.47$ fall in the forward $y^{*}$ interval in the first period but in the backward $y^{*}$ interval in the second period. The systematic uncertainties associated with these $\jpsi$ candidates are fully correlated, assuming they do not depend on the data-taking period. This assumption is checked, and no time dependence in the efficiency corrections is found.

On the other hand, $\jpsi$ events with $|y|<0.47$ always fall in the backward $y^{*}$ interval, and $\jpsi$ candidates with $1.47<|y|<2.4$ always fall in the forward $y^{*}$ interval. The systematic uncertainties associated with these candidates are assumed to be uncorrelated. Based on these considerations, the forward-backward correlation of systematic uncertainties is estimated to be $50\%$. In contrast, for the measurement of $R_{\textrm{FB}}$ as a function of $y^{*}$, the corresponding $y$ intervals do not overlap. Therefore, the systematic uncertainties are assumed to be uncorrelated. A summary of systematic uncertainties in $R_{\textrm{FB}}$ is presented in Table~\ref{table:systematicsRFB}.

\begin{table}[htb]
  \caption{Summary of statistical and systematic uncertainties on the forward-backward ratio $R_{\mathrm{FB}}$ for prompt and nonprompt $J/\psi$. The values are quoted as relative uncertainties (in $\%$) and refer to the range of uncertainties over the specified $p_{\mathrm{T}}$ or $y^{*}$ range.}
    \begin{tabular}{
l@{\extracolsep{10pt}}
c@{\extracolsep{10pt}}
c@{\extracolsep{10pt}}
}
    \hline
\hline                             
Uncertainty & $8<p_{\mathrm{T}}<30$ GeV & $|y^{*} |<1.94$ \\
Stat. prompt  & 3.1--8.9 & 3.8--4.8  \\
Syst. prompt & 6.7--11  &  12--19 \\
Stat. nonprompt  & 5.1--8.4 &6.4--10 \\
Syst. nonprompt & 6.7--11  & 12--19 \\
\hline
\hline
\end{tabular}
\label{table:systematicsRFB}
\end{table}

\begin{figure}[htb]
\centering 
\includegraphics[width=.49\textwidth]{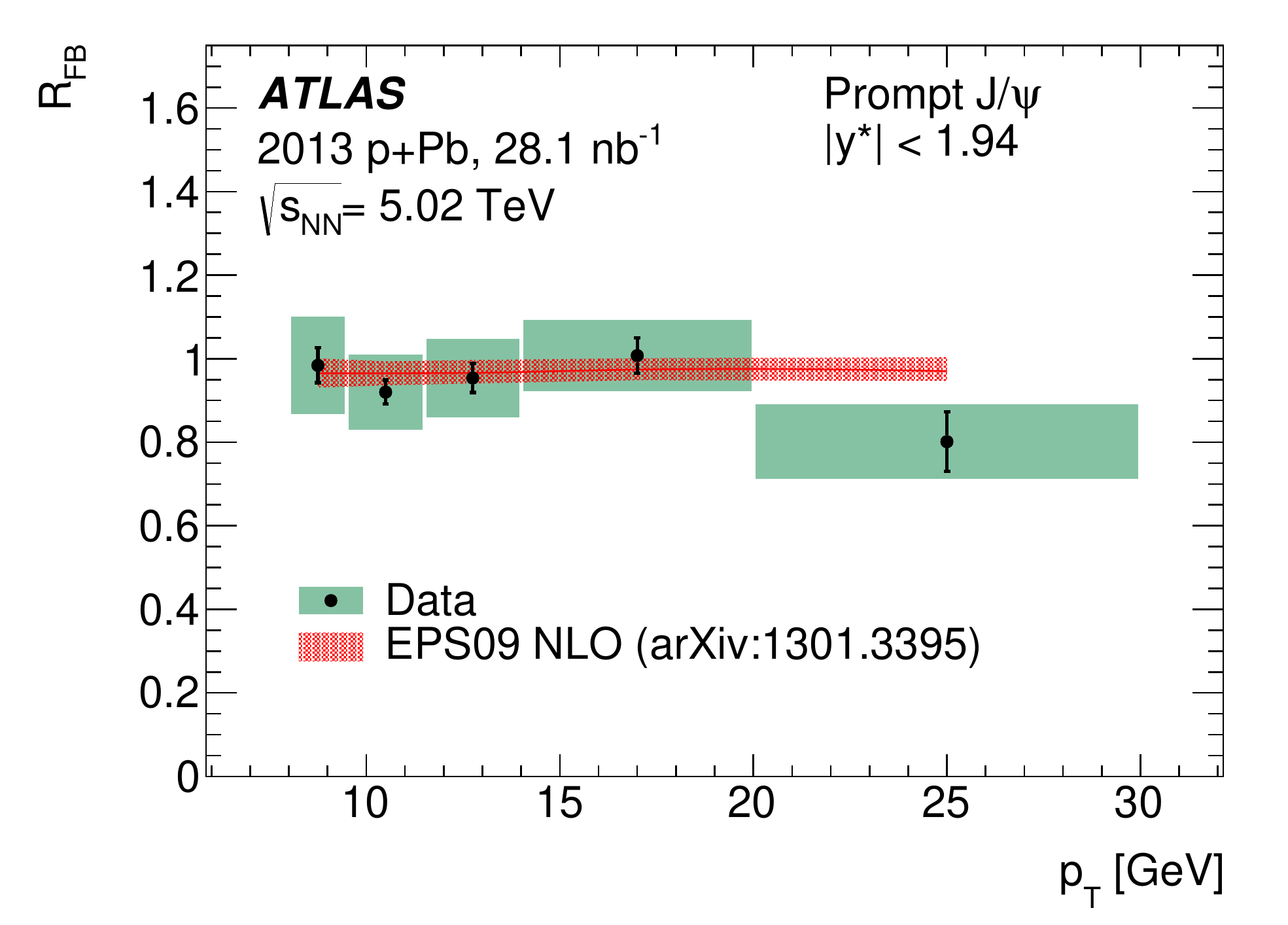}
\includegraphics[width=.49\textwidth]{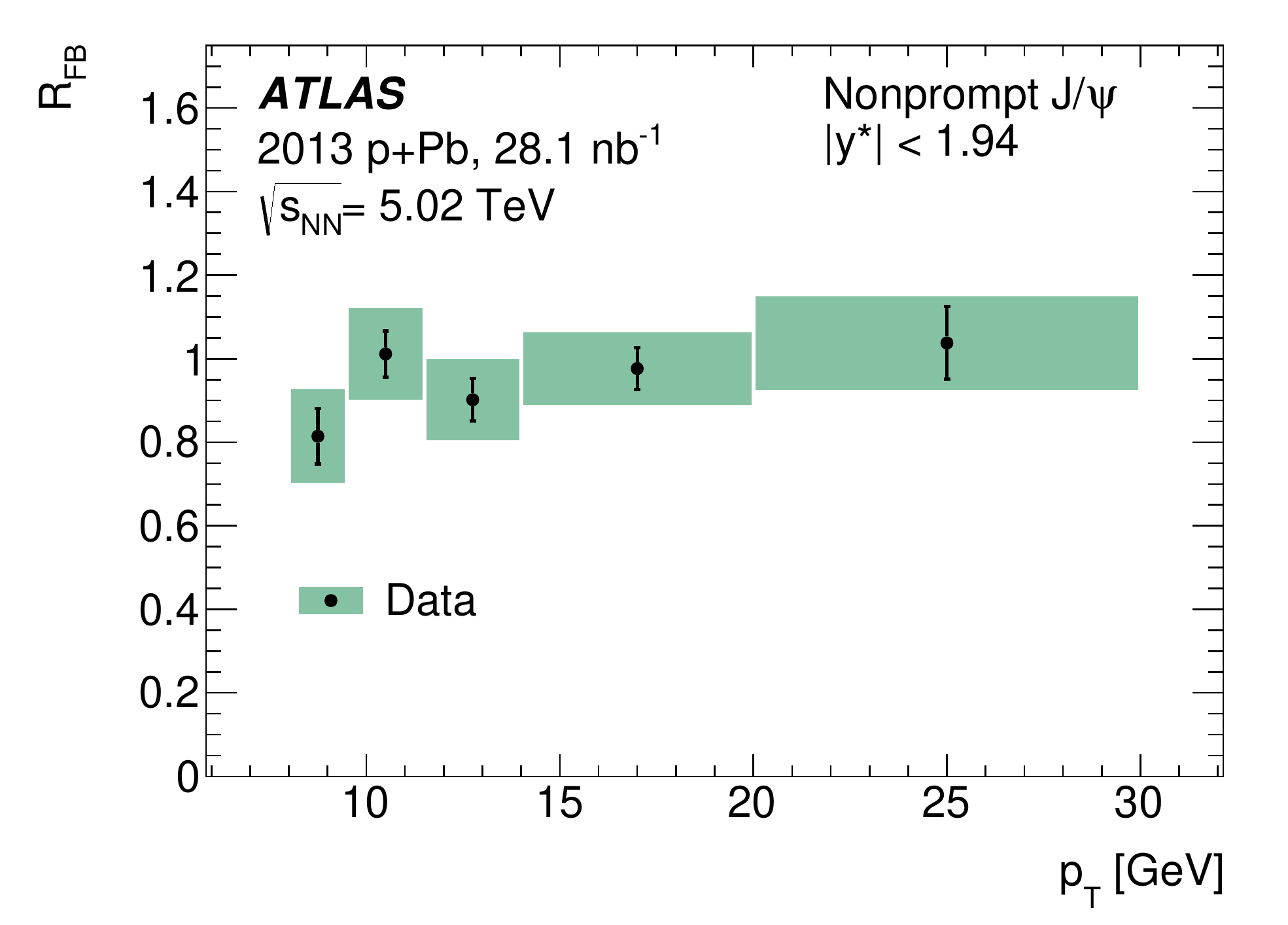}
\caption{Forward-backward production ratio $R_{\mathrm{FB}}$ measured in the center-of-mass rapidity range $|y^{*}| < 1.94$ as a function of $J/\psi$ transverse momentum, for prompt $J/\psi$ (upper panel) and nonprompt $J/\psi$ (bottom panel).
The error bars show the statistical uncertainty, and the shaded boxes show the sum in quadrature of statistical and systematic uncertainties. The narrow horizontal band in the upper panel represents the prediction from Ref.~\cite{Albacete:2013ei} described in the text.}
\label{fig:XsectionRFB}
\end{figure}

\begin{figure}[htb]
\centering 
\includegraphics[width=.49\textwidth]{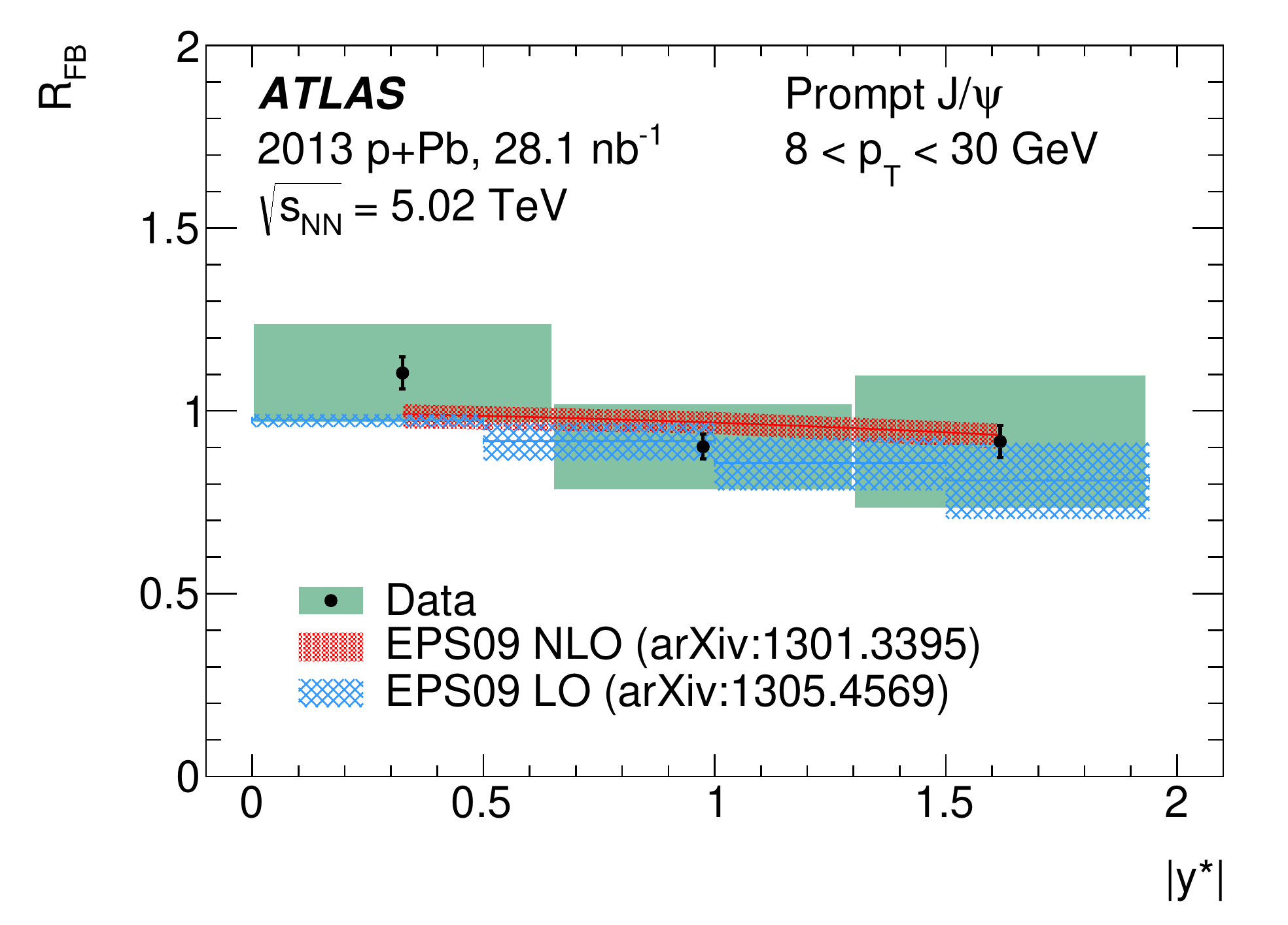}
\includegraphics[width=.49\textwidth]{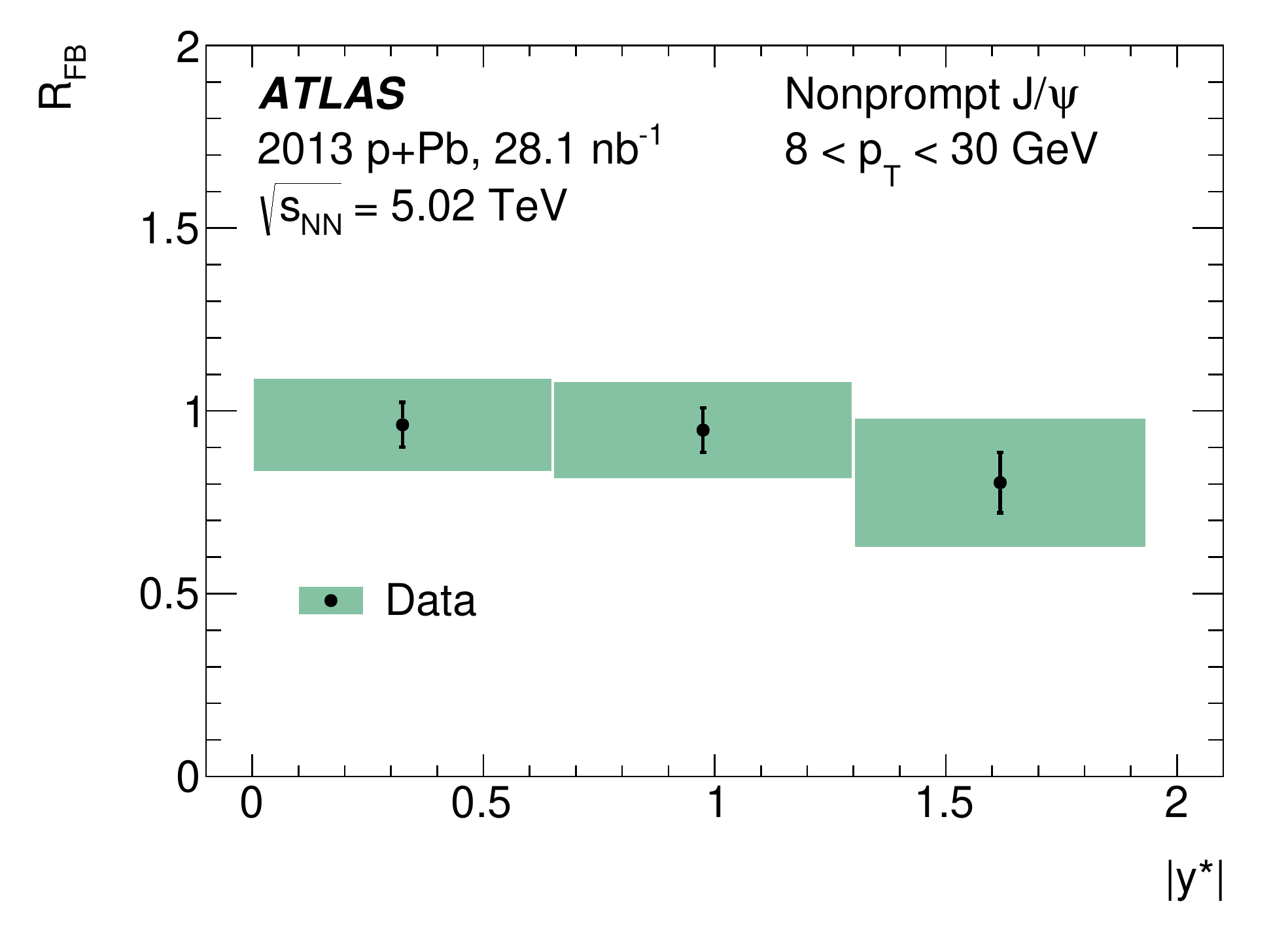}
\caption{Forward-backward production ratio $R_{\mathrm{FB}}$ as a function of center-of-mass rapidity $y^{*}$ for prompt $J/\psi$ (upper panel) and nonprompt $J/\psi$ (bottom panel). The error bars show the statistical uncertainty, and the shaded boxes show the sum in quadrature of statistical and systematic uncertainties. The two bands in the upper panel represent the predictions from Refs.~\cite{Albacete:2013ei,Ferreiro:2013pua} described in the text.}
\label{fig:XsectionRFB_y}
\end{figure}

Figure~\ref{fig:XsectionRFB} shows $R_{\textrm{FB}}$ as a function of transverse momentum in the range $8< \pt < 30\GeV$ for prompt $\jpsi$ (upper panel) and for nonprompt $\jpsi$ (bottom panel). Figure~\ref{fig:XsectionRFB_y} shows $R_{\textrm{FB}}$ as a function of $y^{*}$ in the range $|y^{*}|<1.94$ for prompt $\jpsi$ (upper panel) and for nonprompt $\jpsi$ (bottom panel). These results are consistent with unity within experimental uncertainties. No significant $\pt$ or $y^{*}$ dependence is observed, for both prompt and nonprompt $\jpsi$.

The $R_{\textrm{FB}}$ ratio for prompt $\jpsi$ agrees with theoretical predictions~\cite{Albacete:2013ei,Ferreiro:2013pua} that include shadowing effects based on the EPS09 nuclear parton distribution functions~\cite{Eskola:2009uj}. These results constrain the $y^{*}$ dependence of cold-medium effects in charmonium and $b$-quark production. 


These $R_{\textrm{FB}}$ measurements are complementary to results presented by the LHCb Collaboration, in the range $2.5<|y^{*}|<4.0$, $0<\pt<14\GeV$, that show a difference between prompt and nonprompt $\jpsi$ production, the former showing a strong $\pt$ dependence with values significantly below unity~\cite{Aaij:2013zxa}. The LHCb Collaboration's combined results for inclusive $\jpsi$ production are also consistent with $R_{\textrm{FB}}$ measurements presented by the ALICE Collaboration in the range $2.96<|y^{*}|<3.53$, $0<\pt<15\GeV$~\cite{Abelev:2013yxa}. The difference with respect to the results presented in this paper suggests a strong kinematic dependence of the cold-medium effects on both charmonium and $b$-quark production.

\subsection{Comparison with FONLL calculation}
\begin{figure}[here]
\centering
\includegraphics[width=.48\textwidth]{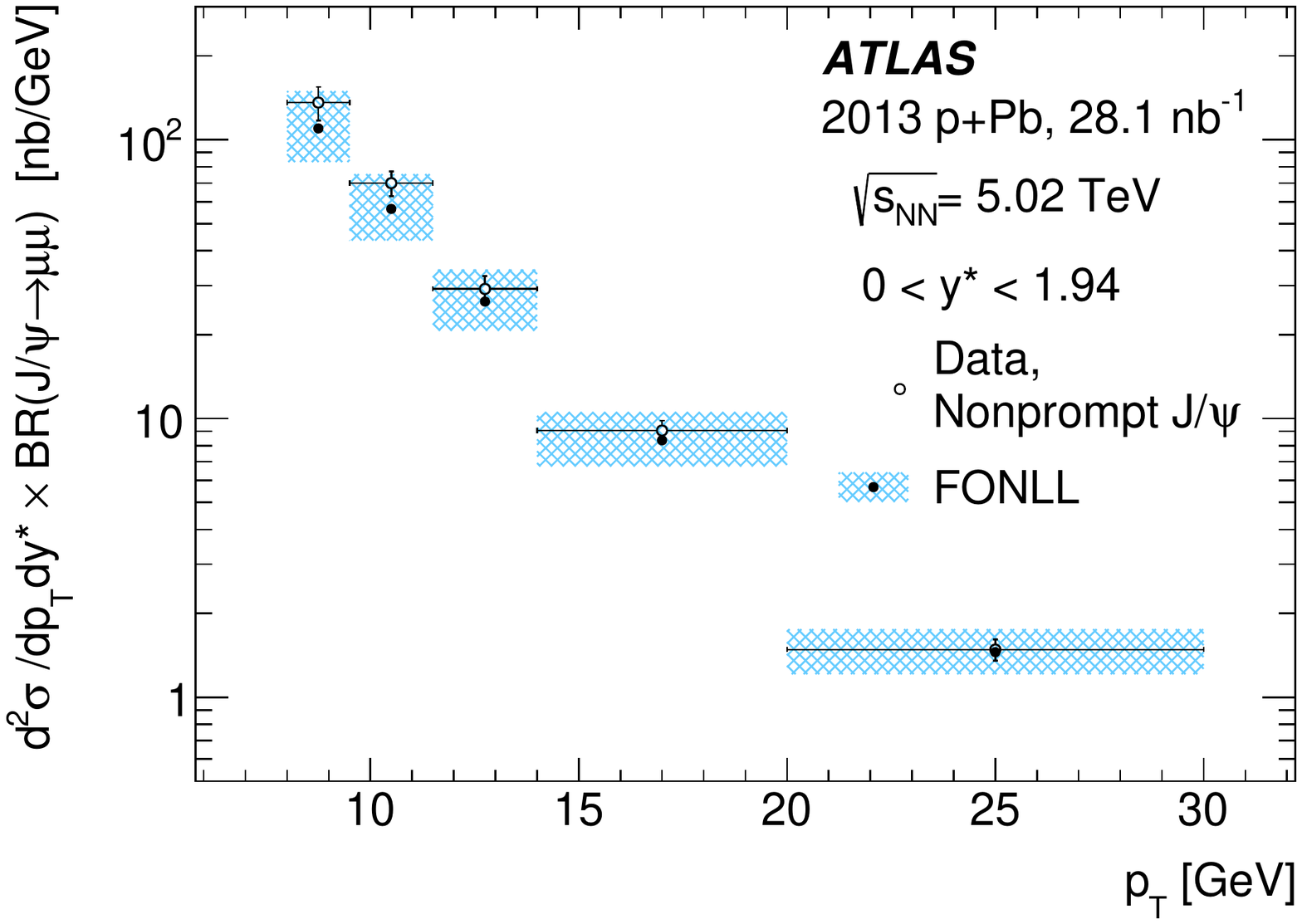}
\includegraphics[width=.48\textwidth]{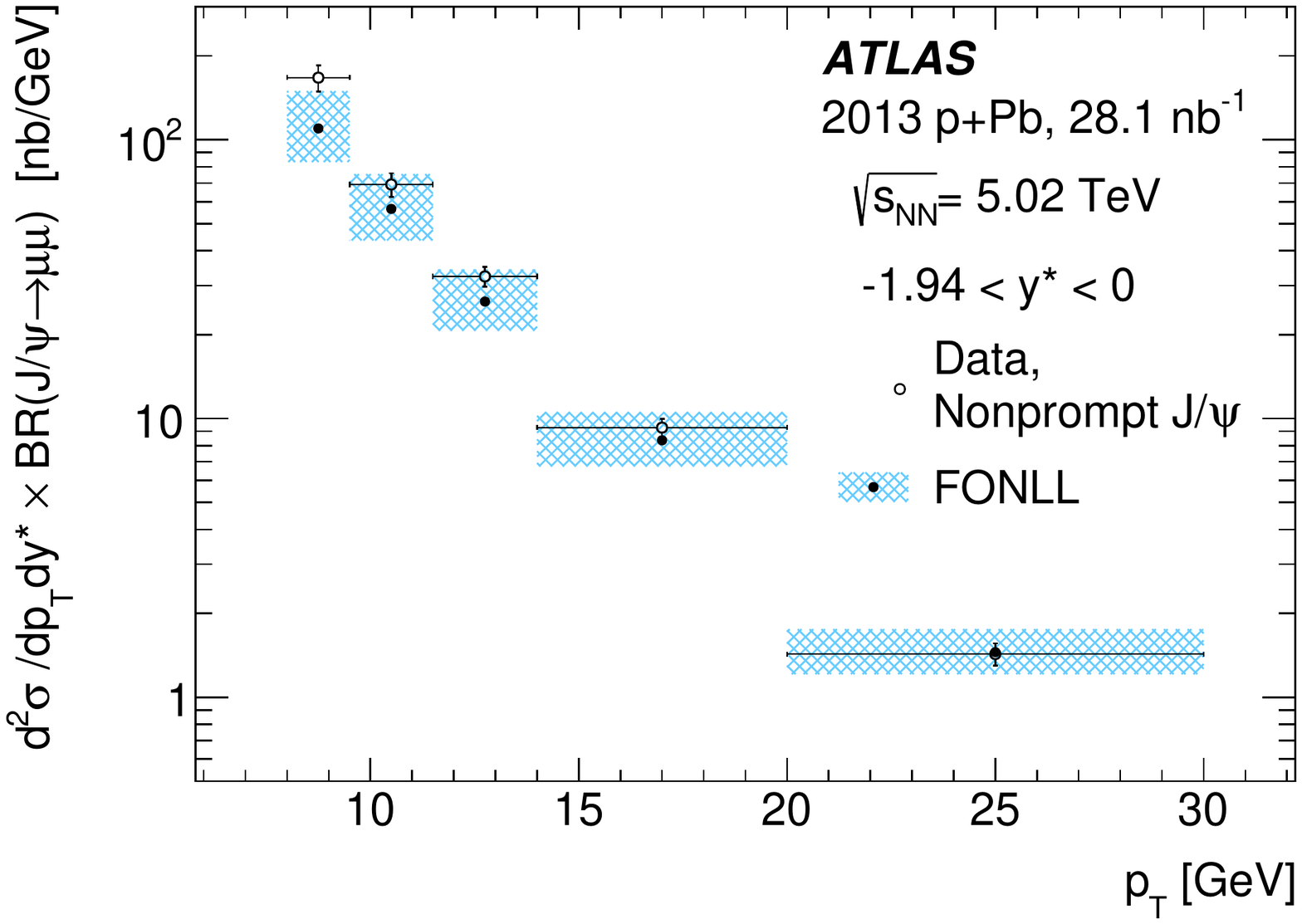}\\
\includegraphics[width=.48\textwidth]{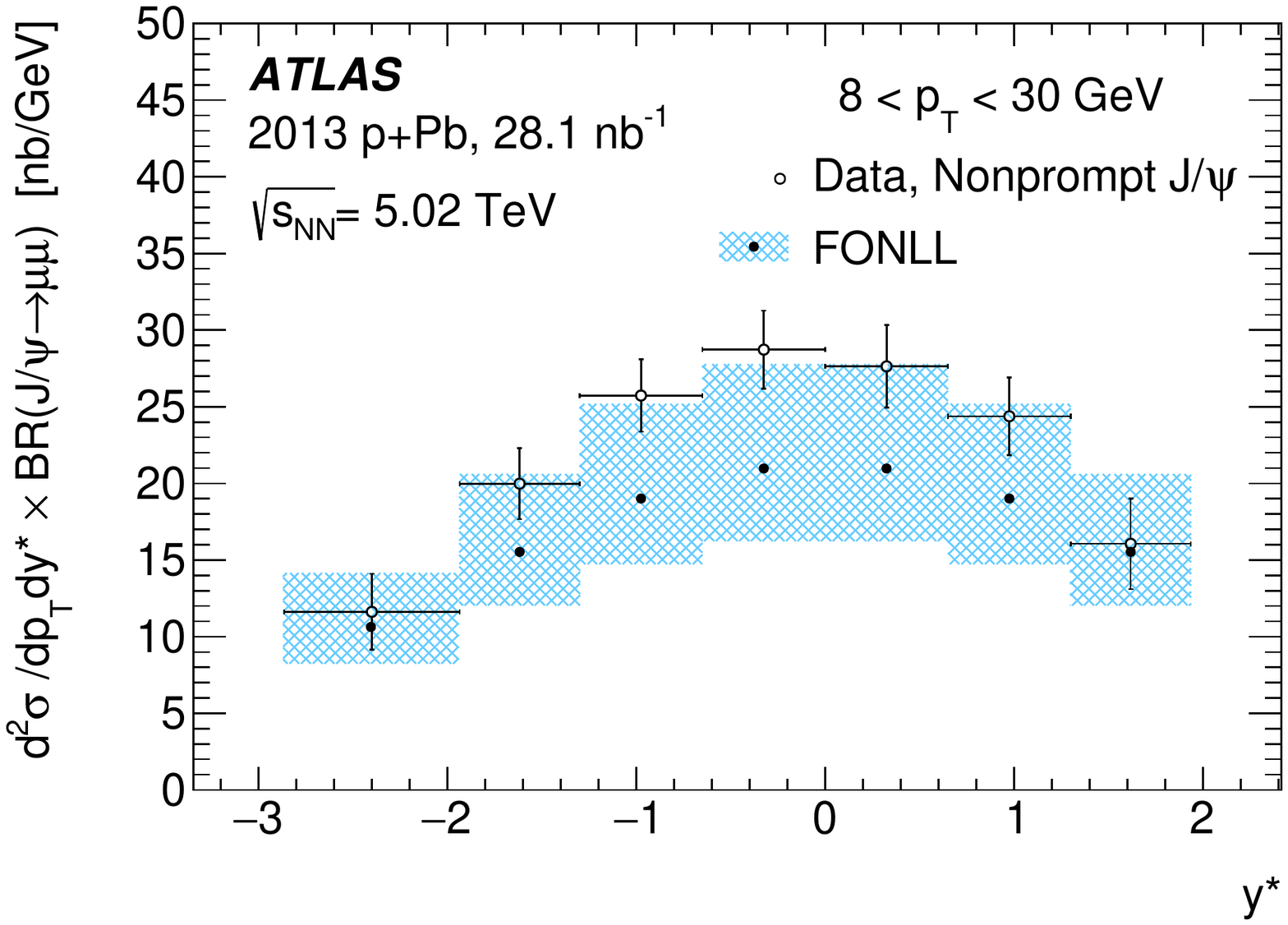}\\
\caption{Differential cross-section for production of nonprompt $J/\psi$ as a function of $J/\psi$ transverse momentum (upper and middle panel) and center-of-mass rapidity (bottom panel) compared with a FONLL calculation for $pp$ collisions scaled by the number of nucleons in the Pb ion. Error bars represent the combination of statistical and systematic uncertainties added in quadrature. The shaded boxes represent the theoretical uncertainties on the FONLL predictions, computed as described in the text. These are strongly correlated between the bins.} 
\label{fig:FONLL}
\end{figure}

The differential cross-sections of nonprompt $\jpsi$ production are compared to FONLL calculations~\cite{Cacciari:2012ny} for $pp$ collisions at 5.02 TeV multiplied by a factor 208 to account for the number of nucleons in the Pb ion. 
The FONLL calculations are performed using CTEQ6.6~\cite{Nadolsky:2008zw} parton distribution functions that do not include any nuclear modification. Systematic uncertainties on the FONLL calculation are obtained by varying the $b$-quark mass (4.75$\pm$0.25$\GeV$), by separately varying the renormalization and factorization scales up and down by a factor of two, and by accounting for parton distribution function uncertainties. As can be seen in Fig.~\ref{fig:FONLL}, the measured cross-sections are consistent with the FONLL calculation within uncertainties.

\FloatBarrier
\section{Conclusions}
In this paper, ATLAS presents measurements of differential cross-sections of prompt and nonprompt $\jpsi$ production in 28.1 $\mathrm{nb}^{-1}$ of $\sqrt{s_{NN}}$ = 5.02$\TeV$ {\em p}+Pb collisions at the LHC in the kinematic range ${-2.87< y^{*}<1.94}$ and ${8<\pt<30\GeV}$. 

The fraction of nonprompt to inclusive $\jpsi$ production is found to depend strongly on $\pt$, reaching values above 50$\%$ at the highest measured $\pt$. No significant $y^{*}$ dependence is observed. This trend is consistent with previous measurements performed with $pp$ data in a similar kinematic range~\cite{Aad:2011sp,Khachatryan:2010yr}. 

The measured differential cross-section for nonprompt $\jpsi$ is compared to a scaled $pp$ reference based on FONLL calculations and is found to be consistent within uncertainties.
   
The measured forward-backward ratios of cross-sections in the range $|y^{*}|<1.94$ are consistent with unity within experimental uncertainties, and with no significant $\pt$ or $y^{*}$ dependence. No difference in these trends is observed between prompt and nonprompt $\jpsi$. These results differ from measurements at more forward $y^{*}$ and lower $\pt$ performed by the LHCb and ALICE Collaborations~\cite{Abelev:2013yxa,Aaij:2013zxa}. This difference suggests a strong kinematic dependence of the cold-medium effects on both charmonium and $b$-quark production..

These results constrain the kinematic dependence of QCD processes in the cold-medium that affect charmonium and $b$-quark production in {\em p}+Pb collisions, and provide a valuable reference for measurements of charmonium and open heavy flavor in Pb+Pb collisions.

\acknowledgments
We thank CERN for the very successful operation of the LHC, as well as the
support staff from our institutions without whom ATLAS could not be
operated efficiently.

We acknowledge the support of ANPCyT, Argentina; YerPhI, Armenia; ARC,
Australia; BMWFW and FWF, Austria; ANAS, Azerbaijan; SSTC, Belarus; CNPq and FAPESP,
Brazil; NSERC, NRC and CFI, Canada; CERN; CONICYT, Chile; CAS, MOST and NSFC,
China; COLCIENCIAS, Colombia; MSMT CR, MPO CR and VSC CR, Czech Republic;
DNRF, DNSRC and Lundbeck Foundation, Denmark; EPLANET, ERC and NSRF, European Union;
IN2P3-CNRS, CEA-DSM/IRFU, France; GNSF, Georgia; BMBF, DFG, HGF, MPG and AvH
Foundation, Germany; GSRT and NSRF, Greece; RGC, Hong Kong SAR, China; ISF, MINERVA, GIF, I-CORE and Benoziyo Center, Israel; INFN, Italy; MEXT and JSPS, Japan; CNRST, Morocco; FOM and NWO, Netherlands; BRF and RCN, Norway; MNiSW and NCN, Poland; GRICES and FCT, Portugal; MNE/IFA, Romania; MES of Russia and NRC KI, Russian Federation; JINR; MSTD,
Serbia; MSSR, Slovakia; ARRS and MIZ\v{S}, Slovenia; DST/NRF, South Africa;
MINECO, Spain; SRC and Wallenberg Foundation, Sweden; SER, SNSF and Cantons of
Bern and Geneva, Switzerland; NSC, Taiwan; TAEK, Turkey; STFC, the Royal
Society and Leverhulme Trust, United Kingdom; DOE and NSF, United States of
America.

The crucial computing support from all WLCG partners is acknowledged
gratefully, in particular from CERN and the ATLAS Tier-1 facilities at
TRIUMF (Canada), NDGF (Denmark, Norway, Sweden), CC-IN2P3 (France),
KIT/GridKA (Germany), INFN-CNAF (Italy), NL-T1 (Netherlands), PIC (Spain),
ASGC (Taiwan), RAL (UK) and BNL (USA) and in the Tier-2 facilities
worldwide.

\providecommand{\noopsort}[1]{}\providecommand{\singleletter}[1]{#1}%

\appendix
\section{Acceptance correction factors}
\label{sec:AppendixA} 
Table~\ref{table:correctionfactor} summarizes the multiplicative correction factors that can be used to correct the central values of $\jpsi$ production cross-sections from isotropic production to an alternative spin-alignment scenario.  The alternative spin-aligment scenarios are described in Ref.~\cite{Aad:2014fpa}.

\section{Tables with results}
The measured $\jpsi$ cross-sections are shown in Table~\ref{table:Prompt} and Table~\ref{table:NonPrompt} for prompt and nonprompt production respectively. The measured nonprompt fractions are shown in 
Table~\ref{table:bfraction}.  The measured forward-backward ratios are shown in Table~\ref{table:RFB}.

\begin{table*}[hbt]
\caption{Scale factors that modify the central cross-section values, evaluated assuming isotropic
decay angular distributions, to a given spin-alignment scenario. The different spin-alignment scenarios
are defined in Ref.~\cite{Aad:2014fpa}.}
\begin{tabular*}{0.5\textwidth}
{lccccc}
 \hline
\hline
$0 <y^{*}<1.94$ &  \multicolumn{5}{c}{$p_{\textrm{T}}$ [GeV]}\\
  &  [8.0,9.5] & [9.5,11.5] & [11.5,14] & [14,20]  & [20,30] \\ 
   Longitudinal & 0.69  &0.70 & 0.71  & 0.74 &0.78 \\
   Transverse zero & 1.29  &1.28 &1.25 &1.22 &1.16 \\
   Transverse positive & 2.79  &1.87 &1.51 &1.36 &1.19 \\
   Transverse negative & 1.02  &1.14  &1.18 &1.17 &1.14 \\
   Off-plane positive & 1.10  &1.11  &1.09 &1.06 &1.04 \\
   Off-plane negative & 0.91 &0.91 &0.93 &0.95 &0.97 \\
\hline
\hline
\end{tabular*}
~\\
~\\
~\\
~\\
\begin{tabular*}{0.5\textwidth}{lccccc}
\hline
\hline
$-1.94 <y^{*}<0$ & \multicolumn{5}{c}{$p_{\textrm{T}}$ [GeV]}\\
   &  [8.0,9.5] & [9.5,11.5] & [11.5,14] & [14,20]  & [20,30] \\
   Longitudinal & 0.68 & 0.69&  0.70&  0.73&  0.78  \\
    Transverse zero & 1.30 & 1.29 &  1.27 &  1.22&  1.16 \\
      Transverse positive & 1.66 &  1.38&   1.30&   1.24& 1.17 \\  
   Transverse negative & 1.10 & 1.22 &  1.23 & 1.21&  1.16 \\  
    Off-plane positive & 1.07 &  1.07&  1.05&  1.03&  1.02 \\ 
    Off-plane negative & 0.94&  0.94&  0.95&  0.97 & 0.98  \\
\hline
\hline
\end{tabular*}
~\\
~\\
~\\
~\\
\begin{tabular}{lccccccc}
\hline
\hline
    $8<p_{\textrm{T}}<30$ GeV & \multicolumn{7}{c}{$y^{*}$} \\ 
     &  [$-$2.87,$-$1.94] &[$-$1.94,$-$1.3] &[$-$1.3,$-$0.65] & [$-$0.65,0] & [0,0.65] & [0.65,1.3] & [1.3,1.94] \\
    Longitudinal & 0.70 & 0.70 &0.69 &0.69 &0.70  &0.70  &0.70 \\
    Transverse zero &1.27 &1.27 &1.28 &1.30 &1.28 &1.26 &1.27 \\
    Transverse positive &3.74 &1.47 &1.47 &1.48 &1.48 &1.49 &4.83 \\ 
    Transverse negative &1.03 &1.14 &1.17 &1.18 &1.15 &1.12 &0.98 \\
    Off-plane positive &1.10 &1.10 &1.06 &1.03 &1.08 &1.11 &1.09 \\
    Off-plane negative &0.91 &0.91 &0.95 &0.98 &0.93 &0.91 &0.92 \\      
\hline
\hline
\end{tabular}
\label{table:correctionfactor}
\end{table*}

\begin{table*}[htb]
\caption{Measured prompt $J/\psi$ differential cross-section multiplied by branching ratio.}
\begin{tabular*}{\textwidth}
{
r@{\extracolsep{0pt}--}
l@{\extracolsep{20pt}}
lr@{\extracolsep{0pt}$\,\pm\,$}
l@{\extracolsep{0pt}$\,\pm\,$}
l@{\extracolsep{0pt}$\,\pm\,$}
l@{\extracolsep{\fill}}
@{\extracolsep{30pt}}
lr@{\extracolsep{0pt}$\,\pm\,$}
l@{\extracolsep{0pt}$\,\pm\,$}
l@{\extracolsep{0pt}$\,\pm\,$}
l@{\extracolsep{\fill}}
}
\hline
\hline
 \multicolumn{3}{c}{} &  \multicolumn{9}{c}{$d^{2}\sigma/dp_{\textrm{T}} dy\times$ BR$(J/\psi\rightarrow\mu\mu)$ [nb/GeV]}  \\
 \multicolumn{2}{c}{$p_{\textrm{T}}$ [GeV]} & & \multicolumn{4}{c}{ $-1.94 < y^{*} < 0$} & &  \multicolumn{4}{c}{ $0 < y^{*} < 1.94$} \\
8.0&9.5& &     414 &      12 (stat) &   39 (syst)  &      11 (lumi) &           & 408 &      12 (stat) &   50 (syst)  &      11 (lumi) \\
9.5&11.5& &     173 &      4 (stat)  &     16 (syst)  &      5 (lumi) &       &   159 &      4 (stat)  &     15 (syst)  &      4 (lumi)\\
11.5&14.0&   &   58.2&      1.4 (stat)  &      4.3 (syst)  &      1.6 (lumi)&      &  55.5&      1.5 (stat)  &      5.7 (syst)  &      1.5 (lumi)\\
14.0&20.0&  &     11.8 &      0.4 (stat)  &      0.8 (syst) &      0.3 (lumi)&      & 11.9 &      0.3 (stat)  &      0.9 (syst) &      0.3 (lumi)\\
20.0&30.0&  &     1.41&      0.08 (stat)  &      0.10 (syst)  &      0.04 (lumi)&       &  1.13&      0.08 (stat)  &      0.07 (syst)  &      0.03 (lumi)\\
 \hline
\hline
\end{tabular*}
~\\
~\\
\begin{tabular*}{1.105\columnwidth}
{
r@{\extracolsep{0pt}$\,,\,$}
l@{\extracolsep{20pt}}
lr@{\extracolsep{0pt}$\,\pm\,$}
l@{\extracolsep{0pt}$\,\pm\,$}
l@{\extracolsep{0pt}$\,\pm\,$}
l@{\extracolsep{\fill}}
@{\extracolsep{30pt}}
lr@{\extracolsep{0pt}$\,\pm\,$}
l@{\extracolsep{0pt}$\,\pm\,$}
l@{\extracolsep{0pt}$\,\pm\,$}
l@{\extracolsep{\fill}}
}

\hline
\hline
 \multicolumn{3}{c}{} &  \multicolumn{4}{c}{$d^{2}\sigma/dp_{\textrm{T}} dy\times$ BR$(J/\psi\rightarrow\mu\mu)$ [nb/GeV]}  \\
\multicolumn{2}{c}{$y^{*}$} &  & \multicolumn{4}{c}{ $8<p_{\textrm{T}}<30$ GeV} \\
$[-$2.87&$-$1.94] & &  43.3   &      1.7 (stat)  &     8.0 (syst)  &      1.2 (lumi) \\
$[-$1.94&$-$1.30] & & 49.0   &      1.3 (stat)  &     5.1 (syst)  &      1.3 (lumi) \\
$[-$1.30&$-$0.65]  &  &58.7   &      1.6 (stat)  &     4.7 (syst)  &      1.6 (lumi) \\
$[-$0.65 & 0.00]  & & 57.1   &      1.7 (stat)  &     4.3 (syst)  &      1.5 (lumi) \\
$[$0.00 & 0.65] & & 63.1 & 1.6 (stat)  &     5.5 (syst)  &      1.7 (lumi) \\
$[$0.65 & 1.30] & &53.0 & 1.4 (stat)  &     5.0 (syst)  &      1.4 (lumi) \\
$[$1.30 & 1.94] & &44.9 & 1.8 (stat)  &     7.2 (syst)  &      1.2 (lumi) \\
\hline
\hline
\end{tabular*}
\label{table:Prompt}
\end{table*}

\begin{table*}[htb]
\caption{Measured nonprompt $J/\psi$ differential cross-section multiplied by branching ratio.}
\begin{tabular*}{\textwidth}
{
r@{\extracolsep{0pt}--}
l@{\extracolsep{20pt}}
lr@{\extracolsep{0pt}$\,\pm\,$}
l@{\extracolsep{0pt}$\,\pm\,$}
l@{\extracolsep{0pt}$\,\pm\,$}
l@{\extracolsep{\fill}}
@{\extracolsep{30pt}}
lr@{\extracolsep{0pt}$\,\pm\,$}
l@{\extracolsep{0pt}$\,\pm\,$}
l@{\extracolsep{0pt}$\,\pm\,$}
l@{\extracolsep{\fill}}
}
\hline
\hline
 \multicolumn{3}{c}{} &  \multicolumn{9}{c}{$d^{2}\sigma/dp_{\textrm{T}} dy\times$ BR$(J/\psi\rightarrow\mu\mu)$ [nb/GeV]}  \\
 \multicolumn{2}{c}{$p_{\textrm{T}}$ [GeV]} & & \multicolumn{4}{c}{ $-1.94 < y^{*} < 0$} & &  \multicolumn{4}{c}{ $0 < y^{*} < 1.94$} \\
8.0&9.5& &     167 &      9 (stat) &   16 (syst)  &      5 (lumi) &           & 136 &      8 (stat) &   17 (syst)  &      4 (lumi) \\
9.5&11.5& &     69.1 &      2.6 (stat)  &     6.3 (syst)  &      1.9 (lumi)  &     &   69.9 &      2.8 (stat)  &     6.6 (syst)  &      1.9 (lumi)\\
11.5&14.0& &   32.3&      1.2 (stat)  &      2.4 (syst)  &      0.9 (lumi)&      &  29.2&      1.3 (stat)  &      3.0 (syst)  &      0.8 (lumi)\\
14.0&20.0& &     9.28 &      0.33 (stat)  &      0.63 (syst) &      0.25 (lumi)&      & 9.06 &      0.33 (stat)  &      0.70 (syst) &      0.24 (lumi)\\
20.0&30.0& &     1.43&      0.08 (stat)  &      0.10 (syst)  &      0.04 (lumi)&       &  1.48&      0.09 (stat)  &      0.09 (syst)  &      0.04 (lumi)\\
\hline
\hline
\end{tabular*}
~\\
~\\
\begin{tabular*}{1.108\columnwidth}
{
r@{\extracolsep{0pt}$\,,\,$}
l@{\extracolsep{20pt}}
lr@{\extracolsep{0pt}$\,\pm\,$}
l@{\extracolsep{0pt}$\,\pm\,$}
l@{\extracolsep{0pt}$\,\pm\,$}
l@{\extracolsep{\fill}}
@{\extracolsep{30pt}}
lr@{\extracolsep{0pt}$\,\pm\,$}
l@{\extracolsep{0pt}$\,\pm\,$}
l@{\extracolsep{0pt}$\,\pm\,$}
l@{\extracolsep{\fill}}
}

\hline
\hline
 \multicolumn{3}{c}{} &  \multicolumn{4}{c}{$d^{2}\sigma/dp_{\textrm{T}} dy\times$ BR$(J/\psi\rightarrow\mu\mu)$ [nb/GeV]}  \\
\multicolumn{2}{c}{$y^{*}$} &  & \multicolumn{4}{c}{ $8<p_{\textrm{T}}<30$ GeV} \\
$[-$2.87&$-$1.94] & &  11.6   &      1.2 (stat)  &     2.2 (syst)  &      0.3 (lumi) \\
$[-$1.94 & $-$1.30]  & & 20.0   &      1.0 (stat)  &     2.1 (syst)  &      0.5 (lumi) \\
$[-$1.30 & $-$0.65]  &  &25.7   &      1.2 (stat)  &     2.0 (syst)  &      0.7 (lumi) \\
$[-$0.65 & 0.00]  & & 28.7   &      1.3 (stat)  &     2.2 (syst)  &      0.8 (lumi) \\
$[$0.00 & 0.65] & &27.6 & 1.2 (stat)  &     2.4 (syst)  &      0.7 (lumi) \\
$[$0.65 & 1.30] & &24.4 & 1.1 (stat)  &     2.3 (syst)  &      0.7 (lumi) \\
$[$1.30 & 1.94] & &16.1 & 1.5 (stat)  &     2.6 (syst)  &      0.4 (lumi) \\
\hline
\hline
\end{tabular*}
\label{table:NonPrompt}
\end{table*}

\begin{table*}[htb]
\caption{Measured fraction of nonprompt $J/\psi$ production.}
\begin{tabular*}{\textwidth}
{
r@{\extracolsep{0pt}--}
l@{\extracolsep{20pt}}
lr@{\extracolsep{0pt}$\,\pm\,$}
l@{\extracolsep{0pt}$\,\pm\,$}
l@{\extracolsep{\fill}}
@{\extracolsep{60pt}}
lr@{\extracolsep{0pt}$\,\pm\,$}
l@{\extracolsep{0pt}$\,\pm\,$}
l@{\extracolsep{\fill}}
}
\hline
\hline
\multicolumn{2}{c}{$p_{\textrm{T}}$ [GeV]} &  & \multicolumn{3}{c}{ $-1.94 < y^{*} < 0$} & &  \multicolumn{3}{c}{ $0 < y^{*} < 1.94$} \\
8.0&9.5& &    0.287 &     0.013 (stat)  &    0.012 (syst)  & &  0.250 &     0.013 (stat)  &    0.023 (syst) \\
9.5&11.5& &     0.286 &   0.009 (stat)  &    0.017 (syst)  & &0.305 &   0.010 (stat)  &    0.020 (syst)  \\
11.5&14.0& &   0.357 &    0.010 (stat)  &   0.015 (syst)  &  &0.345 &    0.012 (stat)  &   0.029 (syst) \\
14.0&20.0& & 0.441 &     0.012 (stat)  &     0.015 (syst)   & & 0.433 &     0.012 (stat)  &     0.022 (syst) \\
20.0&30.0& & 0.504 &      0.021 (stat)  &      0.018 (syst)  & & 0.568 &      0.022 (stat)  &      0.014 (syst) \\
\hline
\hline
\end{tabular*}
~\\
~\\
\begin{tabular*}{1.0\columnwidth}
{
r@{\extracolsep{0pt}$\,,\,$}
l@{\extracolsep{20pt}}
lr@{\extracolsep{0pt}$\,\pm\,$}
l@{\extracolsep{0pt}$\,\pm\,$}
l@{\extracolsep{\fill}}
}
\hline
\hline
\multicolumn{2}{c}{$y^{*}$} &  & \multicolumn{3}{c}{ $8<p_{\textrm{T}}<30$ GeV} \\
$[-$2.87 & $-$1.94] & &  0.212   &      0.019 (stat)  &     0.036 (syst)  \\
$[-$1.94 & $-$1.30]&  &  0.290   &      0.012 (stat)  &     0.023 (syst)  \\
$[-$1.30 & $-$0.65]&  &  0.305   &      0.012 (stat)  &     0.012 (syst)  \\
$[-$0.65 & 0.00] & &  0.335   &      0.013 (stat)  &     0.010 (syst)  \\
$[$0.00 & 0.65]& & 0.305 & 0.011 (stat)  &     0.016 (syst)  \\
$[$0.65 & 1.30]& &  0.315 & 0.012 (stat)  &     0.019 (syst)   \\
$[$1.30 & 1.94]& & 0.264 & 0.019 (stat)  &     0.038 (syst)   \\
\hline
\hline
\end{tabular*}
\label{table:bfraction}
\end{table*}

\begin{table*}[htb]
\caption{Measured forward-backward production ratio.}
\begin{tabular*}{\textwidth}
{
r@{\extracolsep{0pt}--}
l@{\extracolsep{20pt}}
lr@{\extracolsep{0pt}$\,\pm\,$}
l@{\extracolsep{0pt}$\,\pm\,$}
l@{\extracolsep{\fill}}
@{\extracolsep{60pt}}
lr@{\extracolsep{0pt}$\,\pm\,$}
l@{\extracolsep{0pt}$\,\pm\,$}
l@{\extracolsep{\fill}}
}
\hline
\hline
\multicolumn{2}{c}{$y^*$} &  & \multicolumn{3}{c}{ Prompt $J/\psi$} & &  \multicolumn{3}{c}{ Nonprompt $J/\psi$} \\
0.00&0.65&  &   1.10 &     0.04 (stat)  &    0.13 (syst)  & &  0.96 &     0.06 (stat)  &    0.11 (syst) \\
0.65&1.30 & &     0.90 &   0.03 (stat)  &    0.11 (syst)  & &0.95 &   0.06 (stat)  &    0.12 (syst)  \\
1.30&1.94  &   &   0.92 &    0.04 (stat)  &   0.18 (syst)  &  &0.80 &    0.08 (stat)  &   0.15 (syst) \\
\hline
\hline
\end{tabular*}
~\\
~\\
\begin{tabular*}{\textwidth}
{
r@{\extracolsep{0pt}--}
l@{\extracolsep{20pt}}
lr@{\extracolsep{0pt}$\,\pm\,$}
l@{\extracolsep{0pt}$\,\pm\,$}
l@{\extracolsep{\fill}}
@{\extracolsep{60pt}}
lr@{\extracolsep{0pt}$\,\pm\,$}
l@{\extracolsep{0pt}$\,\pm\,$}
l@{\extracolsep{\fill}}
}
\hline
\hline
\multicolumn{2}{c}{$p_{\textrm{T}}$ [GeV]} &  & \multicolumn{3}{c}{  Prompt $J/\psi$} & &  \multicolumn{3}{c}{ Nonprompt $J/\psi$} \\
8.0&9.5& &  0.98 &     0.04 (stat)  &    0.11 (syst)  & &  0.81 &     0.07 (stat)  &    0.09 (syst) \\
9.5&11.5& &     0.92 &   0.03 (stat)  &    0.09 (syst)  & &1.01 &   0.05 (stat)  &    0.09 (syst)  \\
11.5&14.0& &   0.95&    0.03 (stat)  &   0.09 (syst)  &  &0.90 &    0.05 (stat)  &   0.08 (syst) \\
14.0&20.0& & 1.01 &     0.04 (stat)  &     0.07 (syst)   & & 0.98 &     0.05 (stat)  &     0.07 (syst) \\
20.0&30.0& & 0.80 &      0.07 (stat)  &      0.05 (syst)  & & 1.04 &      0.09 (stat)  &      0.07 (syst) \\
\hline
\hline
\end{tabular*}
\label{table:RFB}
\end{table*}

\clearpage
\newpage
\onecolumngrid

\begin{flushleft}
{\Large The ATLAS Collaboration}

\bigskip

G.~Aad$^{\rm 85}$,
B.~Abbott$^{\rm 113}$,
J.~Abdallah$^{\rm 152}$,
S.~Abdel~Khalek$^{\rm 117}$,
O.~Abdinov$^{\rm 11}$,
R.~Aben$^{\rm 107}$,
B.~Abi$^{\rm 114}$,
M.~Abolins$^{\rm 90}$,
O.S.~AbouZeid$^{\rm 159}$,
H.~Abramowicz$^{\rm 154}$
% ...
\bigskip
\\
% ...
$^{11}$ Institute of Physics, Azerbaijan Academy of Sciences, Baku, Azerbaijan\\
% ...
$^{85}$ CPPM, Aix-Marseille Universit{\'e} and CNRS/IN2P3, Marseille, France\\
% ...
$^{90}$ Department of Physics and Astronomy, Michigan State University, East Lansing MI, United States of America\\
% ...
$^{107}$ Nikhef National Institute for Subatomic Physics and University of Amsterdam, Amsterdam, Netherlands\\
% ...
$^{113}$ Homer L. Dodge Department of Physics and Astronomy, University of Oklahoma, Norman OK, United States of America\\
$^{114}$ Department of Physics, Oklahoma State University, Stillwater OK, United States of America\\
% ...
$^{117}$ LAL, Universit{\'e} Paris-Sud and CNRS/IN2P3, Orsay, France\\
% ...
$^{152}$ Institute of Physics, Academia Sinica, Taipei, Taiwan\\
% ...
$^{154}$ Raymond and Beverly Sackler School of Physics and Astronomy, Tel Aviv University, Tel Aviv, Israel\\
% ...
$^{159}$ Department of Physics, University of Toronto, Toronto ON, Canada\\
% ...
$^{*}$ Deceased
\end{flushleft}

\end{document}